\documentclass[conference]{IEEEtran}
\IEEEoverridecommandlockouts
\usepackage{amsmath,amssymb,amsfonts}
\usepackage{algorithmic}
\usepackage{graphicx}
\usepackage{textcomp}
\usepackage{xcolor}
\usepackage{hyperref}
\usepackage{algorithm} 
\usepackage{algorithmic}
\usepackage{cite}
\usepackage{amsmath,amssymb,amsfonts}
\usepackage{algorithmic}
\usepackage{graphicx}
\usepackage{textcomp}
\usepackage{xcolor}
\usepackage{enumitem}
\usepackage{graphicx}
\usepackage{subcaption}
\usepackage{lipsum}
\usepackage{amssymb}
\usepackage{pifont}
\usepackage{subcaption}
\usepackage{graphicx,subcaption}
\usepackage{mathtools, nccmath}
\usepackage{lipsum}
\usepackage{amsmath}
\usepackage{amsfonts}
\usepackage{amssymb}
\usepackage{graphicx}
\usepackage{lipsum}
\usepackage{tabularx}
\usepackage[symbol]{footmisc}
\def\BibTeX{{\rm B\kern-.05em{\sc i\kern-.025em b}\kern-.08em
    T\kern-.1667em\lower.7ex\hbox{E}\kern-.125emX}}
\begin{document}

\title{5GLoR: 5G LAN Orchestration for enterprise IoT applications}

\author{\IEEEauthorblockN{Sandesh Dhawaskar Sathyanarayana, Murugan Sankaradas and Srimat Chakradhar}
\IEEEauthorblockA{
\textit{NEC Laboratories America, Inc}\\
Princeton, NJ 
}
}

\maketitle
\begin{abstract}
5G-LAN is an enterprise local area network (LAN) that leverages 5G technology for wireless connectivity instead of WiFi. 5G technology is unique: it uses network slicing to distinguish customers in the same traffic class using new QoS technologies in the RF domain. This unique ability is not supported by most enterprise LANs, which rely primarily on DiffServ-like technologies that distinguish among traffic classes rather than customers. We first show that this mismatch in QoS between the 5G network and the LAN affects the accuracy of insights from the LAN-resident analytics applications. We systematically analyze the root causes of the QoS mismatch and propose a first-of-a-kind 5G-LAN orchestrator (5GLoR). 5GLoR is a middleware that applications can use to preserve the QoS of their 5G data streams through the enterprise LAN. In most cases, the loss of QoS is not due to the oversubscription of LAN switches but primarily due to the inefficient assignment of 5G data to queues at ingress and egress ports. 5GLoR periodically analyzes the status of these queues, provides suitable DSCP identifiers to the application, and installs relevant switch re-write rules (to change DSCP identifiers between switches) to continuously preserve the QoS of the 5G data through the LAN.

5GLoR improves the RTP frame level delay and inter-frame delay by 212\% and 122\%, respectively, for the WebRTC application. Additionally, with 5GLoR, the accuracy of two example applications (face detection and recognition) improved by 33\%, while the latency was reduced by about 25\%. Our experiments show that the performance (accuracy and latency) of applications on a 5G-LAN performs well with the proposed 5GLoR compared to the same applications on MEC. This is significant because 5G-LAN offers an order of magnitude more computing, networking, and storage resources to the applications than the resource-constrained MEC, and mature enterprise technologies can be used to deploy, manage, and update IoT applications.

\end{abstract}

\section{Introduction}
\begin{figure}[t]
\centerline{\includegraphics[width=1\linewidth]{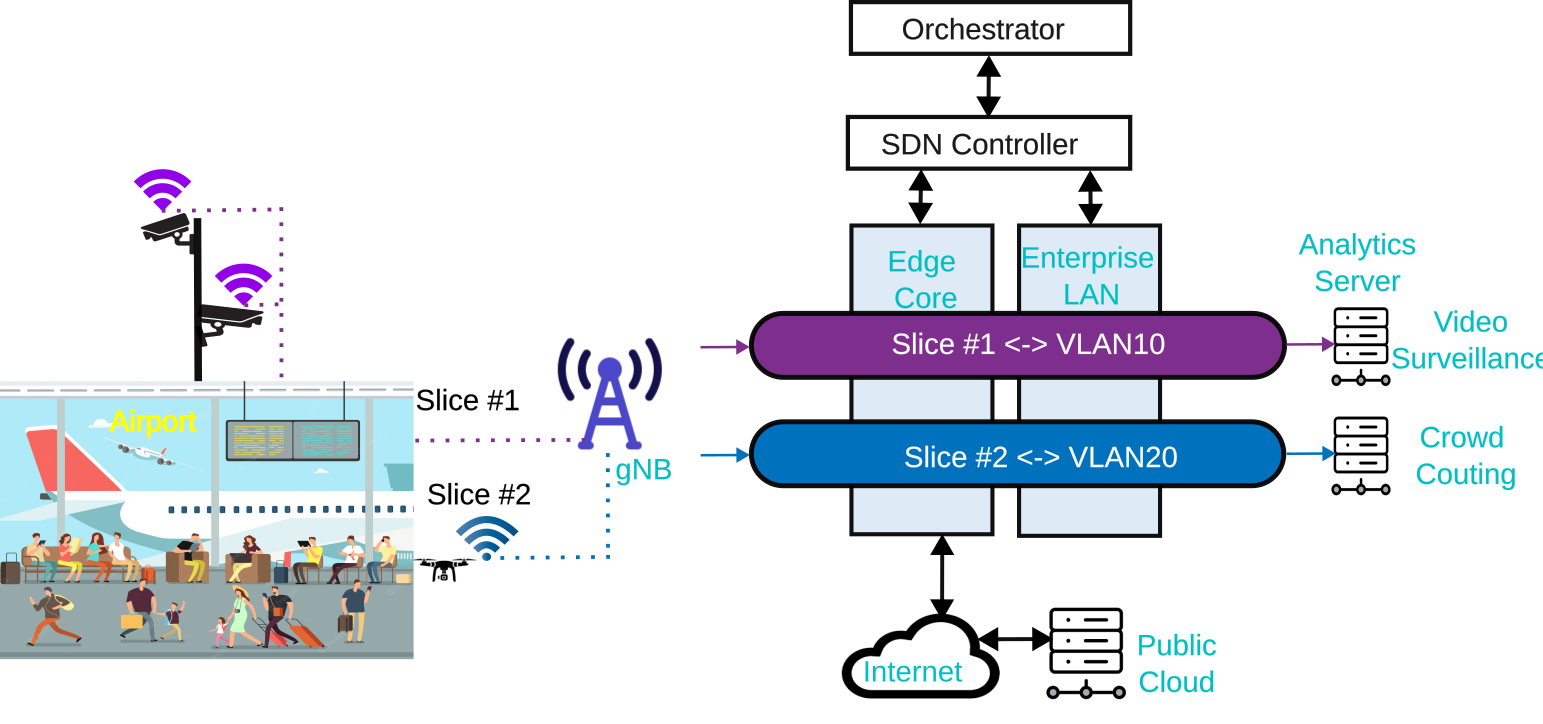}}
\caption{Airport security system using a 5G-LAN} 
\vspace{-0.25 in}
\label{fig:intro-airport}
\end{figure}

\footnote[1]{ \textbf{Published in IEEE Future Networks World Forum-2022}}
WiFi technologies have long dominated wireless connectivity in the enterprise, but that is poised to change. 5G-LAN is an enterprise local area network (LAN) that leverages 5G technology for wireless connectivity instead of WiFi. A 5G-LAN enables enterprises to benefit from the much wider coverage and higher performance, reliability, and security of 5G cellular technology (compared to WiFi). A 5G-LAN is specifically deployed to meet the high-performance needs of various enterprises by combining traditional enterprise LANs with the power and performance of 5G cellular.
5G-LAN enables new IoT applications in enterprises like factories, warehouses, airports, hospitals, and arenas. One such example is shown in Figure~\ref{fig:intro-airport}, where numerous surveillance video cameras in an airport use a private 5G cellular network (we will refer to it as RAN) to stream high-definition videos to analytics applications like video surveillance, which are hosted on the enterprise LAN of the airport.

The increasing use of private RAN is fueled by the rapid increase in the use of IoT devices in enterprises. According to Ericson, 5 billion IoT devices will be connected over the RAN \cite{erricson} by 2025, and the global 5G IoT market is expected to be worth US \$40 billion by 2026 \cite{5g-market}. AWS already offers private 5G in several regions, including the United States East (Ohio), the United States East (North Virginia), and the United States West (Oregon)\cite{aws5g}.

\begin{figure*}[t]
    \centering
    \begin{subfigure}{0.65\columnwidth}
        \centering
        \includegraphics[width=\columnwidth,height=1.3 in]{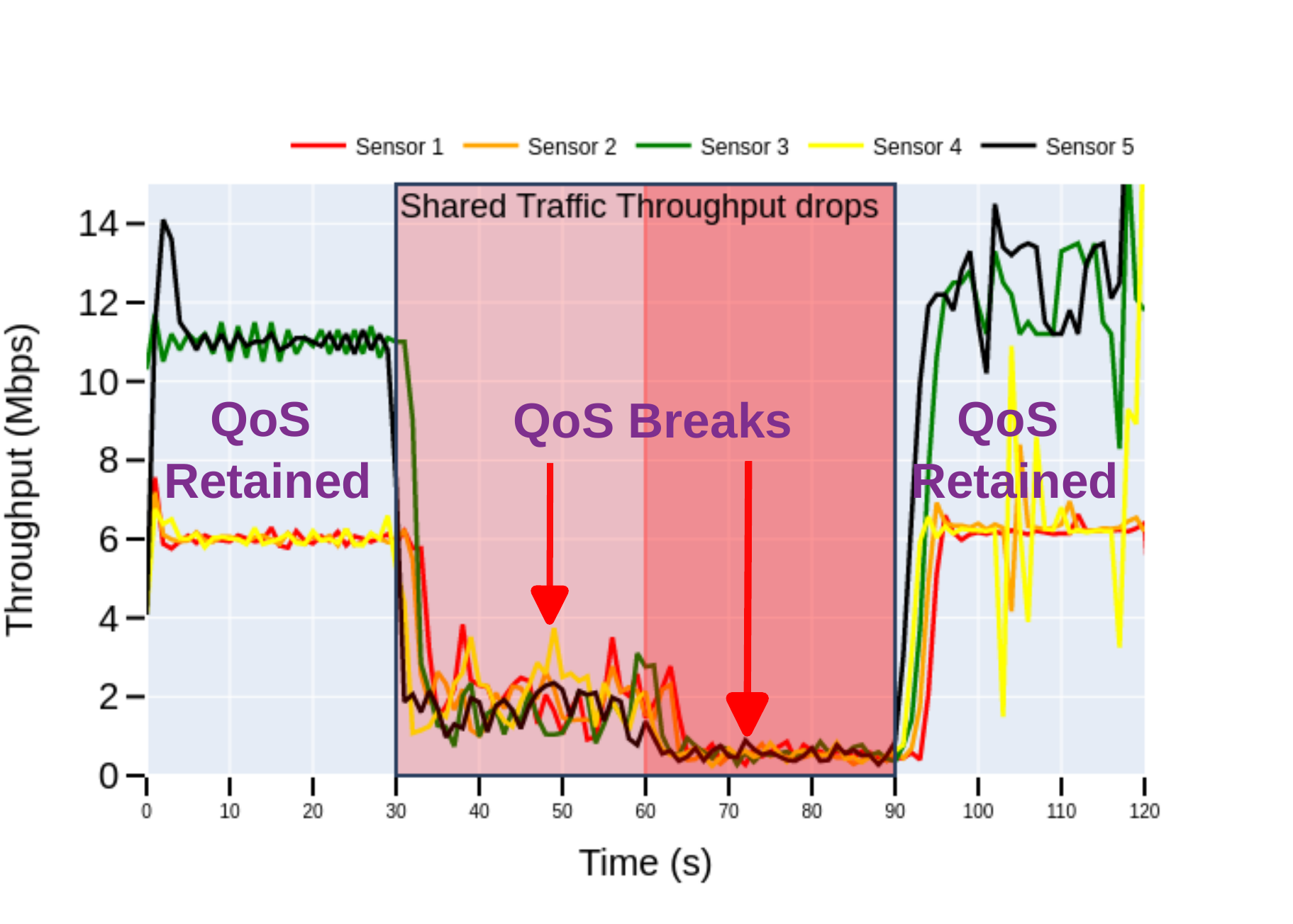}
        \vspace{-0.25 in}
        \caption{Throughput}\label{fig:motivation-tcp}
    \end{subfigure}
    \begin{subfigure}{0.65\columnwidth}
        \centering
        \includegraphics[width=\columnwidth,height=1.3 in]{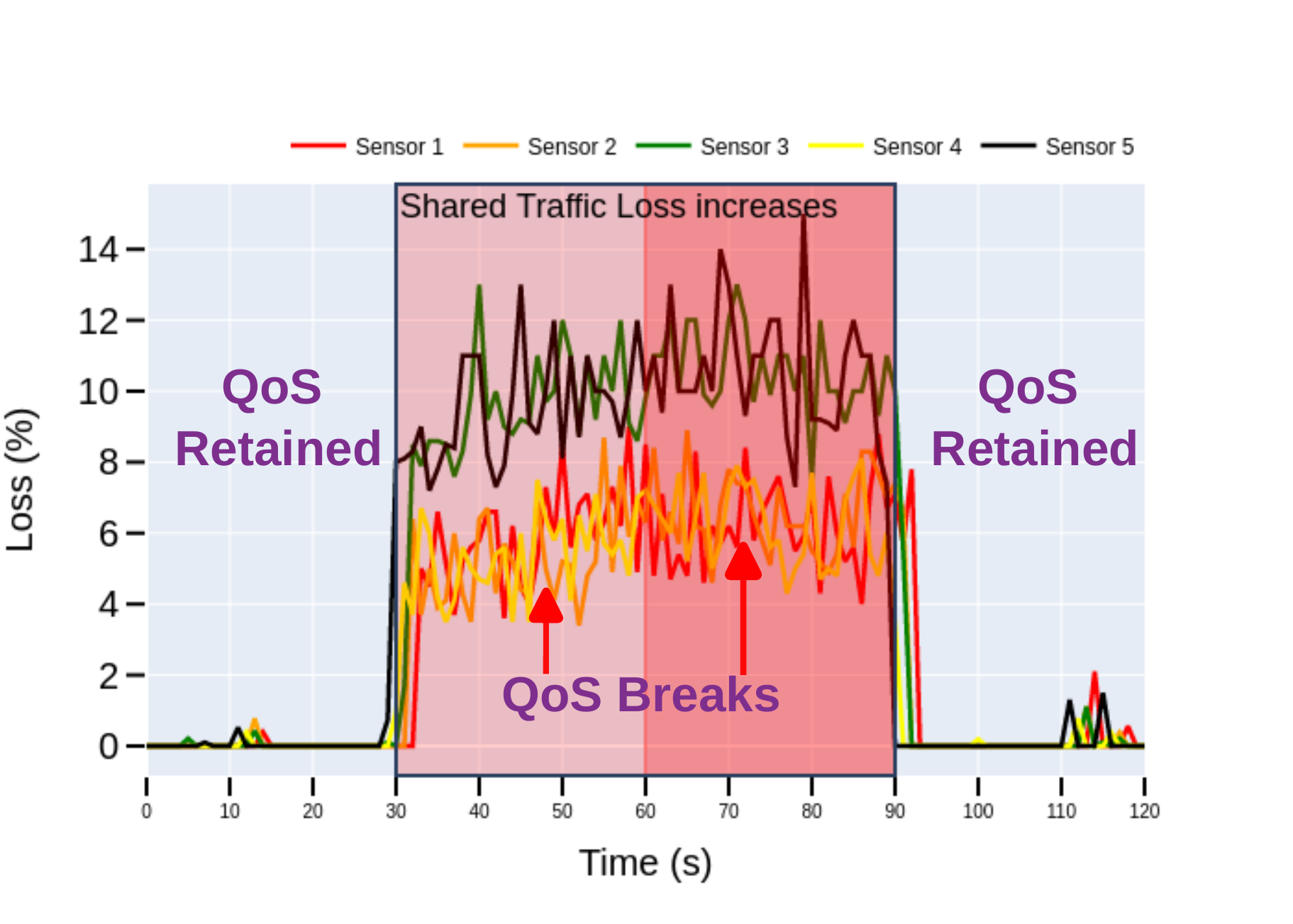}
        \vspace{-0.25 in}
        \caption{Loss}\label{fig:motivation-udp}
    \end{subfigure}
    \begin{subfigure}{0.65\columnwidth}
        \centering
        \includegraphics[width=\columnwidth,height=1.3 in]{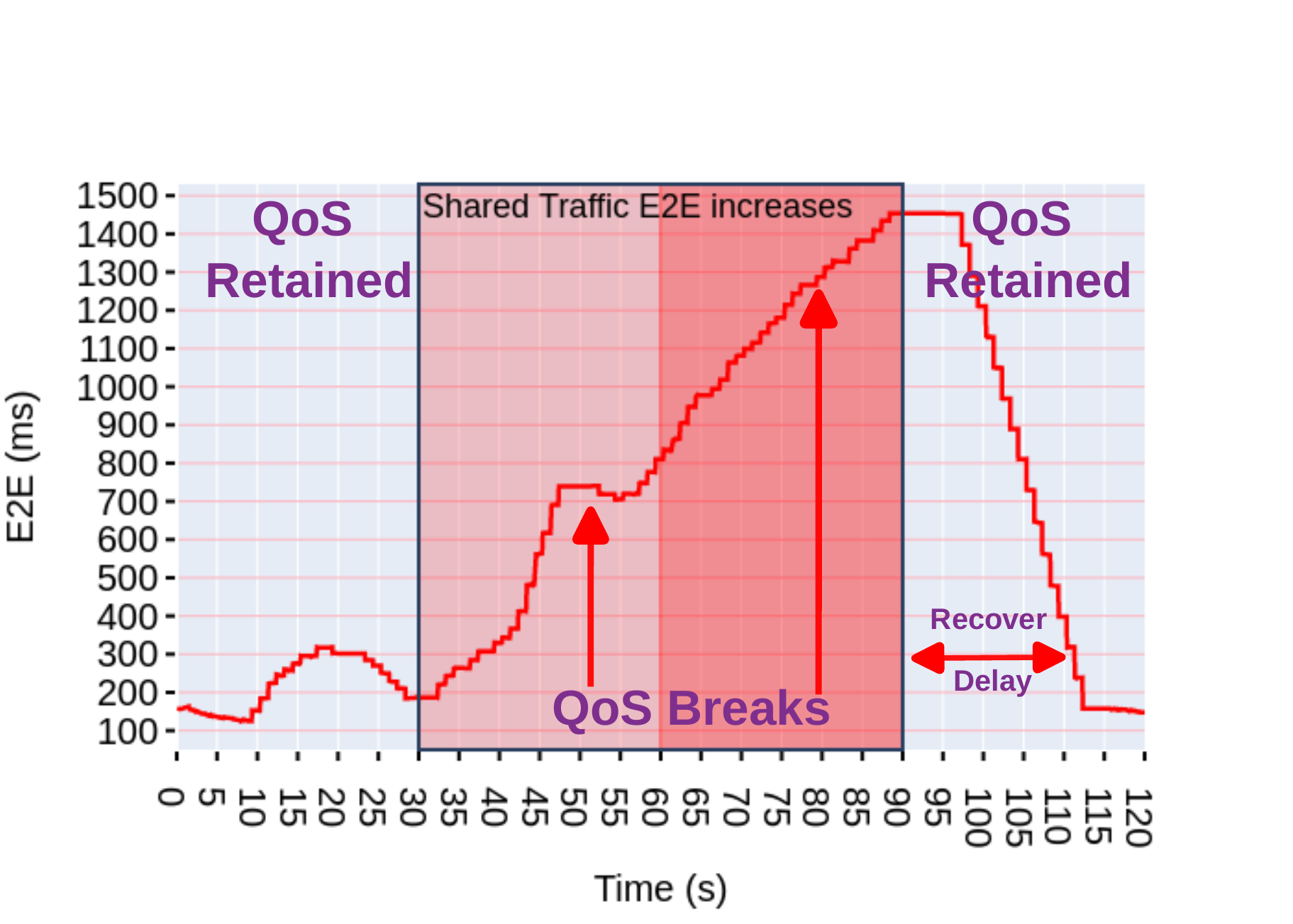}
        \vspace{-0.25 in}
        \caption{Frame Level E2E delay}\label{fig:motivation-webrtc}
    \end{subfigure}
    \vspace{-0.1 in}
    \caption{QoS Analysis of the RAN slice in LAN}
    \label{fig:motivation}
    \vspace{-0.2 in}
\end{figure*}
Cellular operators have provided different quality of service (QoS) for specific traffic classes like video, voice, or data for many years by using technologies like DiffServ in the transport and core network. However, unlike any prior cellular technology, 5G's network slicing technology also supports QoS in the radio frequency (RF) domain. Unlike DiffServ, 5G slicing can provide differential traffic treatment from different customers in the same traffic class, like voice or video, by using new QoS technologies in the RF domain. Consequently, 5G slicing can isolate traffic from different customers in the same traffic class and provide differentiated services to each customer.

For example, consider the airport deployment shown in Figure~\ref{fig:intro-airport}. Surveillance video cameras in the airport deliver video streams to video surveillance applications hosted on the enterprise LAN. Similarly, drones monitoring the outdoor areas of the airport deliver video streams to crowd detection applications that are hosted on the enterprise LAN. Crowd detection applications monitor people or vehicles in outdoor areas of the airport. As shown in Figure~\ref{fig:intro-airport}, surveillance and drone video streams traverse the RAN in two separate slices (slice \#1 and slice \#2). Although the two data streams belong to the same traffic class (video), applying different QoS policies (like priority or security) to the two data streams is possible. 

This unprecedented ability of 5G slicing to differentially treat customers in the same traffic class is not supported by most enterprise LANs. LAN is primarily based on DiffServ-like technologies that provide varying QoS for different traffic classes rather than different customers. Therefore, 5G slice customer-specified QoS for data streams (within the same traffic class) ends within the 5G network. Once the streams exit the 5G network, the enterprise LAN cannot ensure customer-specified QoS. We show in later sections that this mismatch in QoS between the cellular and wired networks can adversely affect the accuracy of insights from the analytics applications on the enterprise LAN.

In this paper, we systematically analyze the impact of the mismatch in QoS capabilities between the private RAN and the enterprise LAN. The critical analytics applications on the enterprise LANs are deployed on real-world private 5G deployment. We propose a new application-specific method {\it 5GLoR} (5G-LAN Orchestrator) that aims to preserve the QoS of 5G data streams as they traverse the enterprise LAN.

5GLoR is a real-time, application-specific orchestrator middleware that an application can use to manage its 5G data streams on enterprise LANs. For each 5G data stream in a slice, 5GLoR is aware of the customer-specified QoS requirement (in a 5G slice). 5GLoR first determines appropriate queues via enterprise switches that can maintain the QoS requirement for the 5G data stream. Then, 5GLoR identifies an appropriate DSCP identifier for the 5G data stream based on available queues. The application assigns this identifier to all IP packets in the data stream. As the DSCP-tagged data stream traverses the switches in the LAN, it may be necessary to change the DSCP identifier between switches. 5GLoR installs switch re-write rules to re-write the DSCP identifier, as necessary.

\begin{figure*}[t]
    \centering
    \begin{subfigure}{0.65\columnwidth}
        \centering
        \includegraphics[width=\columnwidth,height=1.3 in]{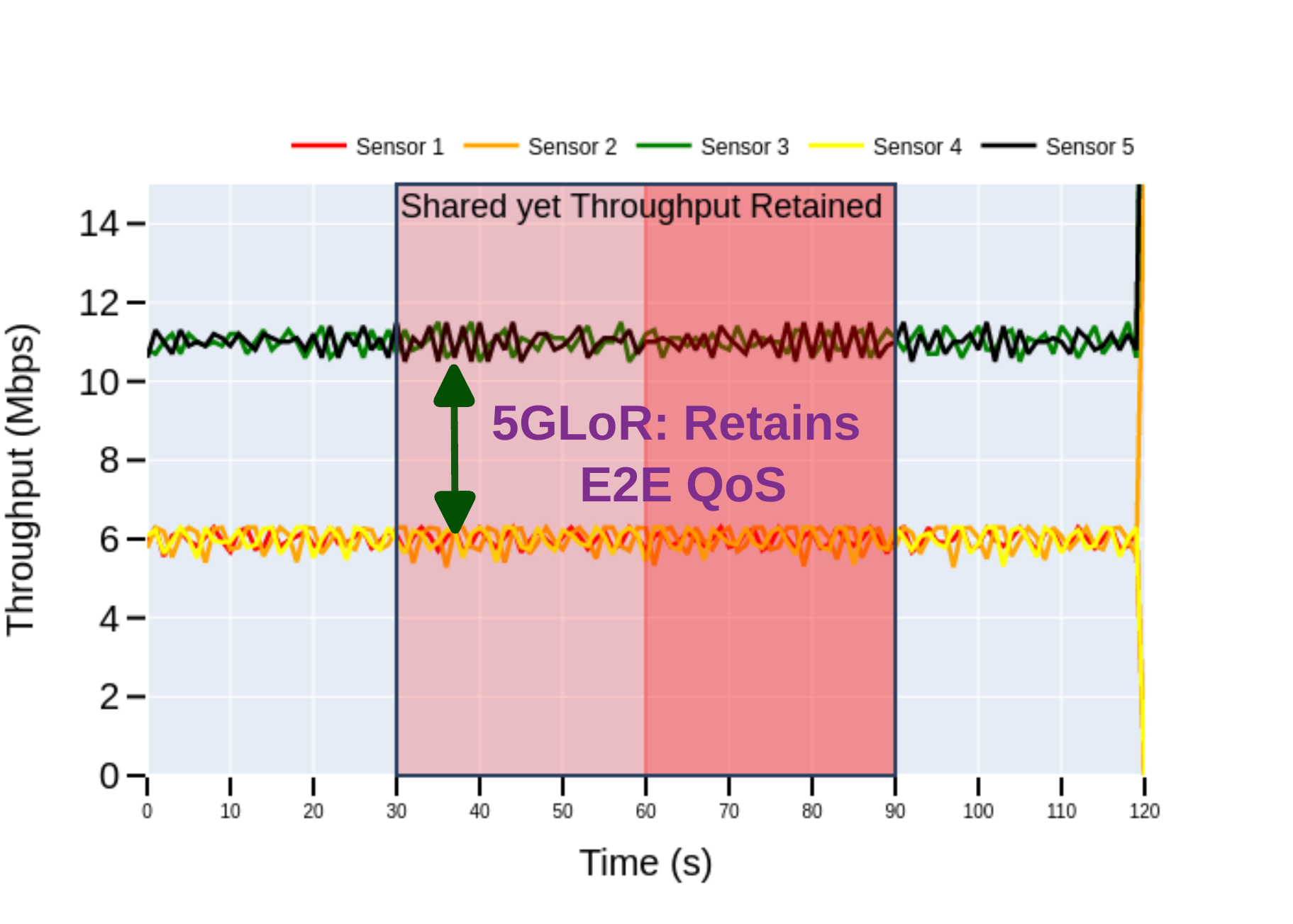}
        \vspace{-0.25 in}
        \caption{Throughput}\label{fig:motivation-tcp-sol}
    \end{subfigure} 
    \begin{subfigure}{0.65\columnwidth}
        \centering
        \includegraphics[width=\columnwidth,height=1.3 in]{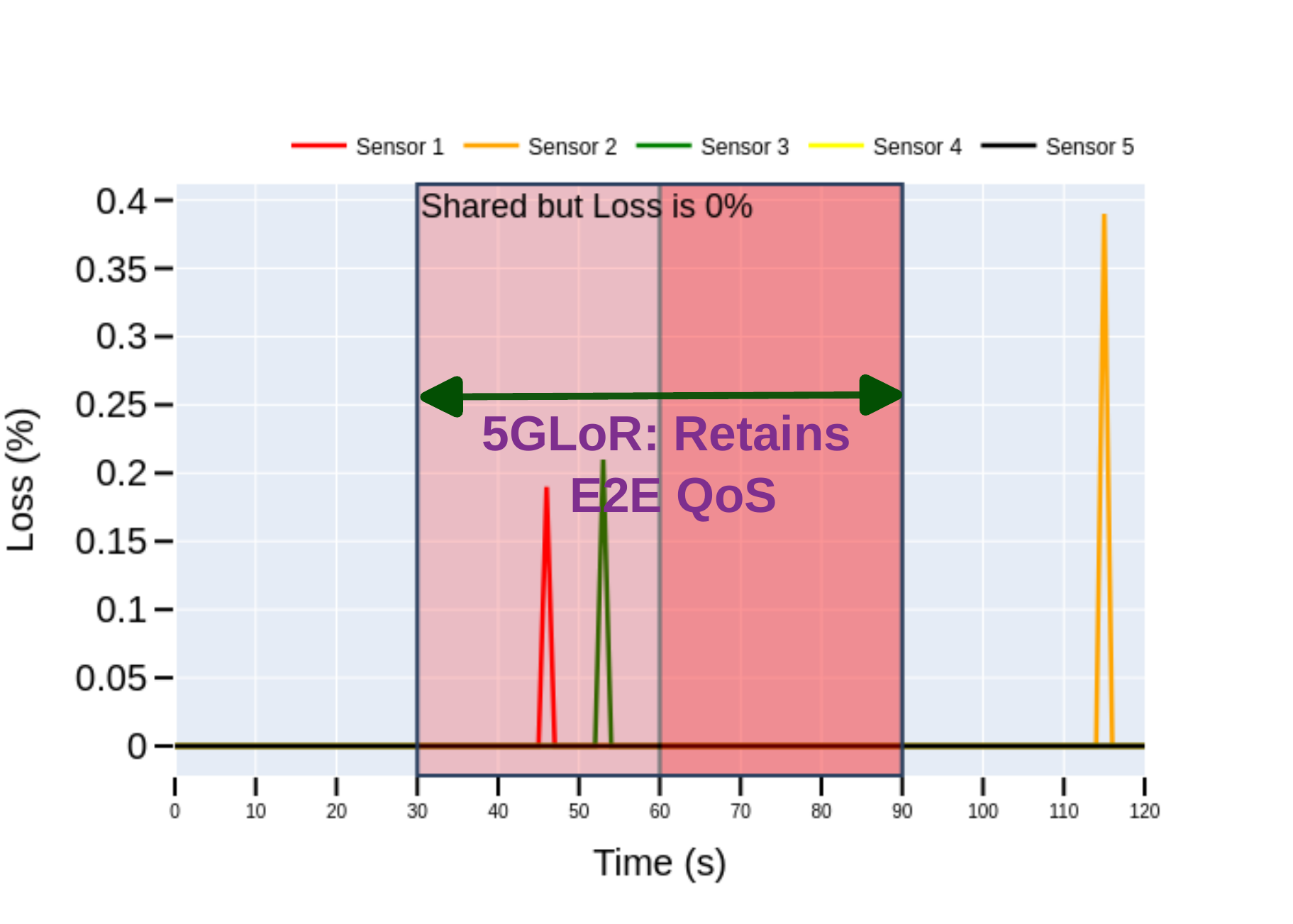}
        \vspace{-0.25 in}
        \caption{Loss}\label{fig:motivation-udp-sol}
    \end{subfigure}
    \begin{subfigure}{0.65\columnwidth}
        \centering
        \includegraphics[width=\columnwidth,height=1.3 in]{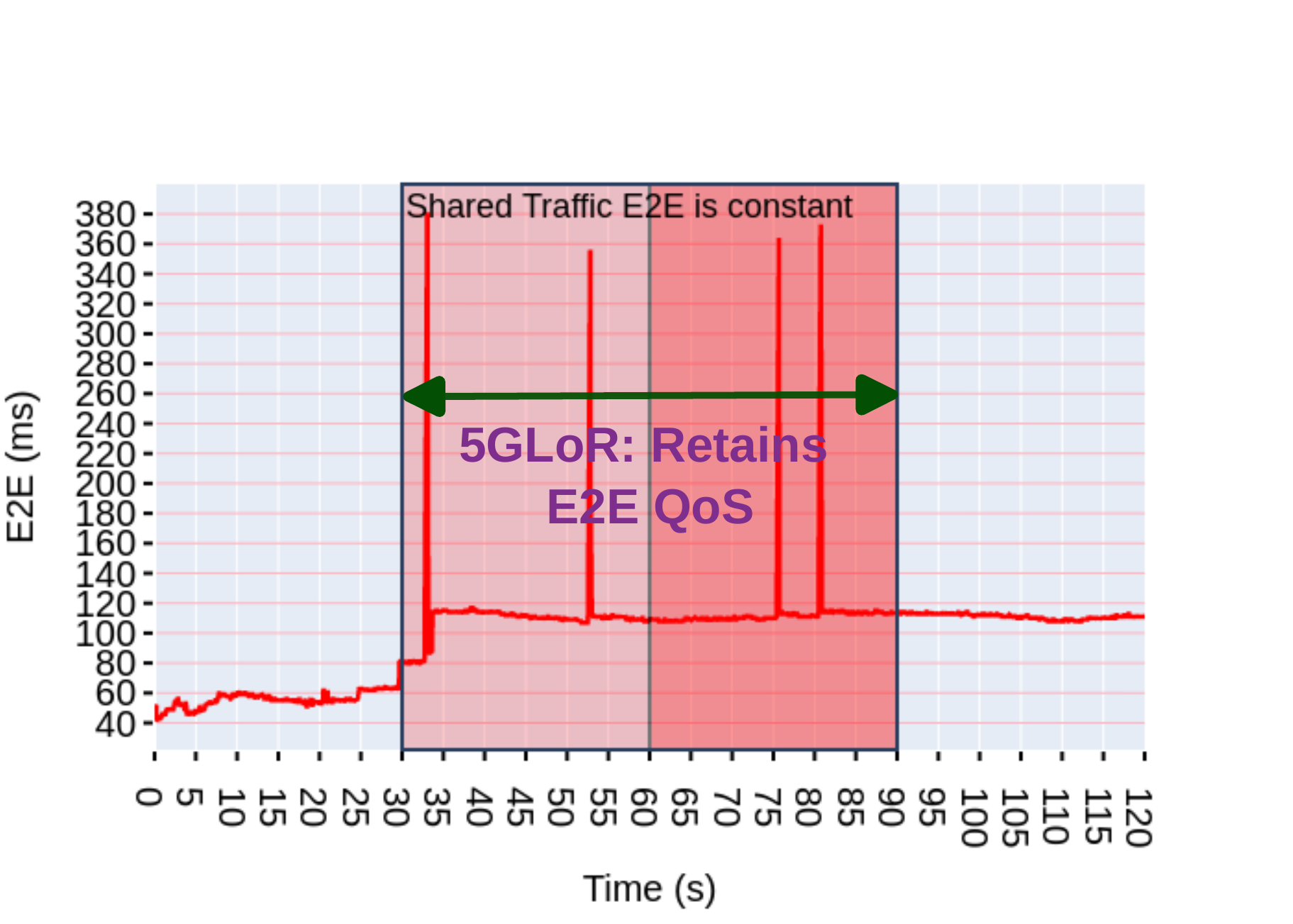}
        \vspace{-0.25 in}
        \caption{Frame Level E2E delay}\label{fig:motivation-webrtc-sol}
    \end{subfigure}
    \vspace{-0.05 in}
    \caption{QoS Analysis for the RAN slice to LAN with proposed solution}
    \label{fig:motivation-sol}
    \vspace{-0.2 in}
\end{figure*}

Additionally, it ensures that the RAN flows do not suffer from local LAN flows with higher priority. Once a DSCP identifier is assigned to a 5G data stream and the switch re-write rules are used, the 5G data stream traverses the LAN with the customer-specified QoS. Our unique approach does not affect the existing non-5G traffic using the enterprise LAN because 5GLoR uses admission control to allow 5G data streams into the enterprise LAN. If 5GLoR cannot find a suitable path to preserve QoS, then 5GLoR allows the 5G data stream to traverse the LAN using a best-effort policy. 


Compared to a 5G-LAN without our proposed orchestration, 5GLoR improves the RTP frame level delay and inter-frame delay by 212\% and 122\%, respectively for WebRTC \cite{WebRTC}. The accuracy of two example applications (face detection and recognition) improved by 33\% and the average time by 25\% when using 5GLoR. We also compared the 5G-LAN with 5GLoR to an edge-computing architecture where all analytics applications are run at the edge of the 5G network (i.e., on the multi-access edge computing or MEC) rather than being hosted on the enterprise LAN. Our experiments show that the performance (accuracy and latency) of applications on a 5G-LAN with the proposed 5GLoR compares well with the performance of the same applications on MEC. This is significant because 5G-LAN offers an order of magnitude more computing, networking, and storage resources to the applications than the resource-constrained MEC, and mature enterprise technologies can be used to deploy, manage, and update IoT applications on enterprise LANs.

In summary, we make the following contributions:
\begin{itemize}
    \item We designed a real-world 5G-LAN composed of a CBRS-based RAN and an enterprise LAN; several real-world video analytics applications like face recognition were hosted on the enterprise LAN, with several video cameras on the RAN. 
    \item We conducted extensive experiments to show the mismatch in QoS capabilities between the slicing-based RAN and the enterprise LAN.
    \item We systematically analyzed the root causes of the QoS mismatch in the enterprise LAN and proposed a novel, application-specific 5G-LAN orchestrator (5GLoR) that preserves the QoS of a 5G data stream as it traverses the enterprise LAN. 
    \item We demonstrated that the addition of 5GLoR to a 5G-LAN significantly improves the performance of three popular video analytics applications, like Face Detection, Face Recognition, and WebRTC. 
    \item We also showed that hosting applications on a 5G-LAN with 5GLoR is competitive with the alternative of hosting applications on MEC at the edge of the RAN.
\end{itemize}{}

The rest of the paper is organized as follows. We discuss the motivation for our work by reporting results for popular video analytics applications on a  real-world private RAN and enterprise LAN. Then, we discuss prior work and how it relates to the proposed 5G-LAN orchestrator 5GLoR. After discussing the system overview and major components of 5GLoR, we present the detailed design of 5GLoR, including design challenges and solutions. Finally, we report a detailed evaluation of 5GLoR in a real-world 5G-LAN setting and compare it with an alternative architecture where applications run on MEC in 5G.

\section{Motivation}\label{sec:motivation}

We demonstrate the impact of the QoS mismatch between the 5G network and the enterprise LAN. We use a real-world 5G-LAN (airport security system described in Section~\ref{sec:eval}), where the application on the enterprise LAN processes videos from five video cameras on the 5G network. These video streams are in five different 5G slices. The five slices have a Guaranteed Bit Rate (GBR) of \{6, 6, 11, 6, 11\} Mbps, and all the flows have a DSCP marking of 39. A slice guarantees QoS (i.e., the RAN reserves the GBR requested). We use DSCP \cite{dscp} identifiers on the LAN to distinguish among traffic classes like video, voice, or data. We show that the QoS guarantee in a slice is not preserved as the 5G data traverses the LAN, adversely impacting the accuracy of insights from the analytics applications hosted on the LAN.

The total test duration is 120 seconds, with four windows of 30 seconds each. During the first window, the LAN does not have any traffic other than the five RAN flows. During the second window, 75 LAN flows with a DSCP marking of 39 are introduced into the LAN, and during the third window, an additional 75 LAN flows are introduced but without any DSCP marking. In the fourth window, we remove all the 150 LAN flows. We measure the application's throughput, packet loss, and end-to-end (E2E) delay in receiving the frames from the video cameras.

During the first window, as shown in fig \ref{fig:motivation-tcp} and \ref{fig:motivation-udp}, the application sees a throughput of 100\% with 0\% packet loss. However, as the 75 DSCP marked LAN flows are introduced during the second window, the application throughput drops by ~81\%, and the application experiences a packet loss of about ~10\%. During the third window, the additional LAN flows (with no DSCP marking) do result in some degradation of the RAN flows (compared to the case in the first window). During the fourth window, there are no LAN flows, and the RAN flows begin to exhibit the same QoS that was observed in the first window. A similar observation is made for the E2E delay of the video frame. As we see in the fig \ref{fig:motivation-webrtc}, during the first window, the E2E delay is small and increases up to 300 ms, and during the second window, it increases to 750 ms (2.3 times the delay in the first window). In the third window, the delay goes up further to 1450ms (4.8 times the delay in the first window). Such large E2E values cause frame drops for real-time video streaming. After all LAN flows are removed in the fourth window, we observe a delay of 300 ms (the same as the delay in the first window). However, it took almost 20 seconds to recover from the increased delay observed in the third window.

It is evident from the above experiments that 5G Slice's GBR (guaranteed bit rate) is preserved on the LAN when there is no other LAN traffic. However, when new LAN traffic is introduced (practical scenario), the QoS guarantee for RAN flows is broken even when the RAN flows have a DSCP marking. This is because of the mismatch in QoS between the 5G network and LAN.

In this paper, we propose 5GLoR to preserve the QoS of RAN flows as they transit through the LAN. When we run the same experiments with the 5GLoR assisting the application, there is no throughput degradation, as shown in Figure~\ref{fig:motivation-tcp-sol}. Also, the packet loss is almost zero, and the video frame (end-to-end) delay is around 380ms (see Figure~\ref{fig:motivation-webrtc-sol}) compared to about 1500ms delay when 5GLoR is not used. More details about the 75 background LAN flows are described in Section~\ref{sec:lanflows}.


\section{Related Work}


\textbf{5G Slice}: The 5G slice construct begins from the wireless mobile device on RAN and ends at the core of the 5G. Much of the previous work has been done on the design of slice in the air (RAN). A.Banchs et al. in \cite{related-1} focus on analyzing two categories of resource allocation, dedicated and shared. V.Sciancalepore et al. in  \cite{related-2} developed ONETS, a theoretical model that emphasizes maximizing the network slicing multiplexing gains rather than providing dedicated or shared resources. 
A.Gohar et al. in \cite{related-3} also build a slice broker that works on minimizing the cost by purchasing the slices from the service provider. All these works emphasize the resource allocation and management of slices from the wireless side, but 5GLoR addresses the issue of QoS breakdown of 5G slice in LAN.


\textbf{5G-LAN}: Celona offers the 5G-LAN infrastructure \cite{b6}, that provides benefits of remote updates to end user devices, better management and  device operation via remote monitoring. However, 5G-LAN, as described above, suffers due to localized traffic in LAN. 5GLoR is built to address this issue.

\textbf{Protocols:} Multiprotocol Label Switching (MPLS) \cite{mpls} and Resource reservation protocol (RSVP)\cite{rsvp} are routing protocols to connect enterprise sites through public internet service providers. These protocols also offer an end-to-end QoS mechanism by creating private tunnels or specific flows. One would ask, why not use these protocols with 5G-LAN? Firstly RSVP is not supported in all the routers, and MPLS is the only class-based QoS guarantee, not flow-based. These are WAN-specific router protocols, and using them in an enterprise LAN adds significant overhead. Hence we build 5GLoR



\section{System Overview of 5GLoR}\label{sec:overview}


\begin{figure}[!t]
\centerline{\includegraphics[width=1\linewidth]{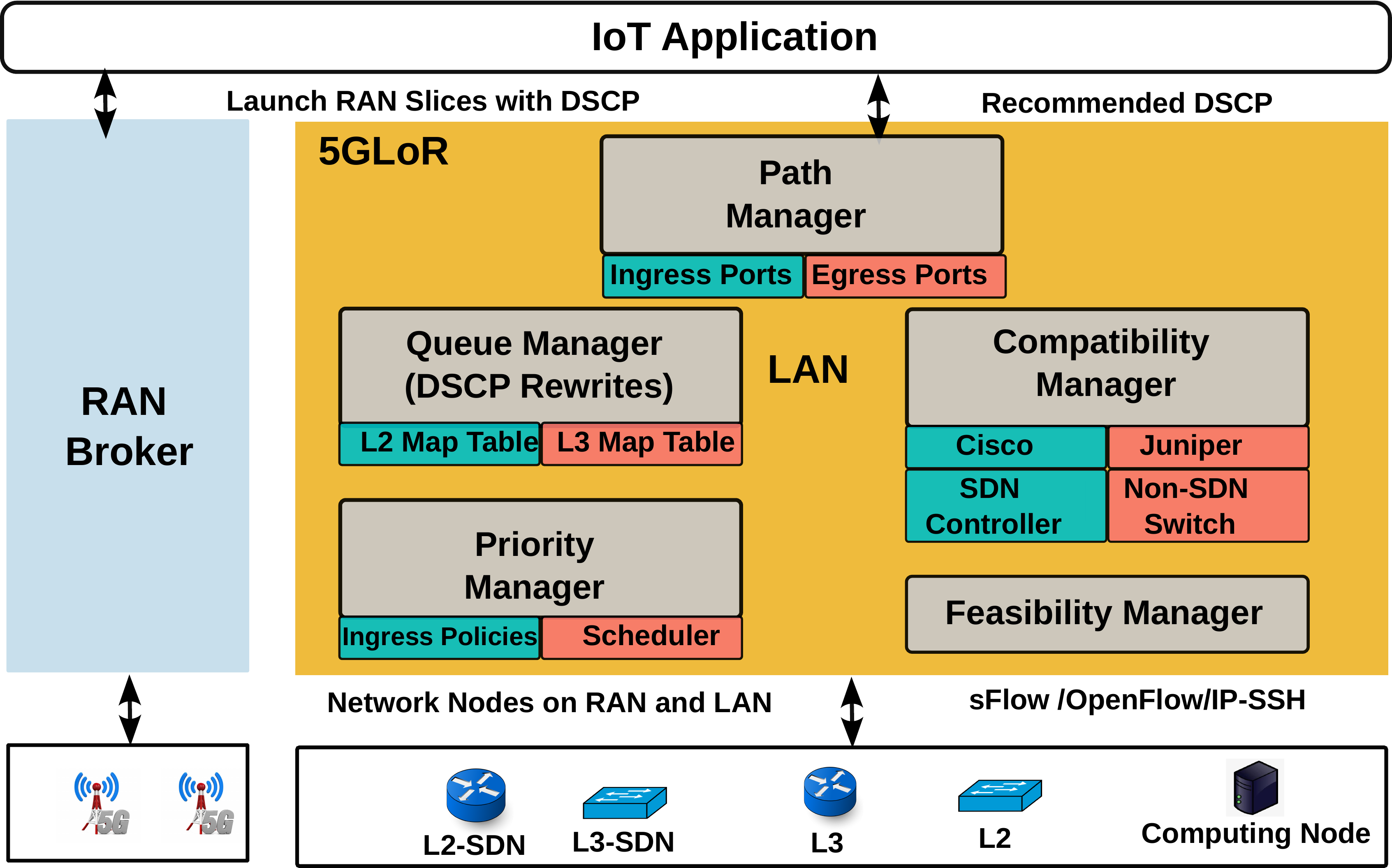}}
\caption{System Overview of 5GLoR}
\label{fig:sys}
\vspace{-0.2 in}
\end{figure}


This section gives a brief overview of 5GLoR.
Key components of 5GLoR are shown in Figure~\ref{fig:sys}. 
The Path Manager determines the path (i.e., switches and the ingress and egress ports on each switch) the 5G data stream will take in the enterprise LAN. This path can also be manually specified. After knowing the path, 5GLoR analyzes the status of the various queues associated with the egress ports of each switch in the path. Such information can be obtained from an SDN (software-defined networking) controller on the LAN or by directly querying the switch. 

Based on the status of the various queues in each switch, the Feasibility Manager determines if there are queues in each switch so that the QoS of the 5G data can be preserved as the data traverses the LAN. If such a path exists, then the Compatibility Manager checks if a single, common DSCP identifier is sufficient to ensure that the QoS of the 5G data is preserved as the data traverses through the various queues across the switches. If a common DSCP identifier across all switches is not available, then 5GLoR uses the Queue Manager to install the appropriate re-write rules on each switch (this will allow the DSCP identifier to be changed mid-way during the path). 5GLoR also periodically monitors the path, and if necessary, it can change the DSCP identifier the application should use or update re-write rules on switches. 
 
If the Feasibility Manager fails to find a suitable path that can preserve the QoS, then 5GLoR alerts the application that Slice QoS will be broken on the LAN and pinpoints the switch and port where the QoS is likely to be broken. If the application wishes to continue to use the LAN, then 5GLoR will enable the 5G data to traverse the offending switch using the best-effort policy. In the following sections, we will provide more details about the goals, challenges, and detailed design of key components in 5GLoR.


\section{Design}

In this section, we discuss the root causes of the QoS mismatch, list the key challenges in designing a solution like 5GLoR, and describe the key components of 5GLoR.

\subsection{Root causes for QoS mismatch}
\begin{figure}[!t]
\centerline{\includegraphics[width=1\linewidth]{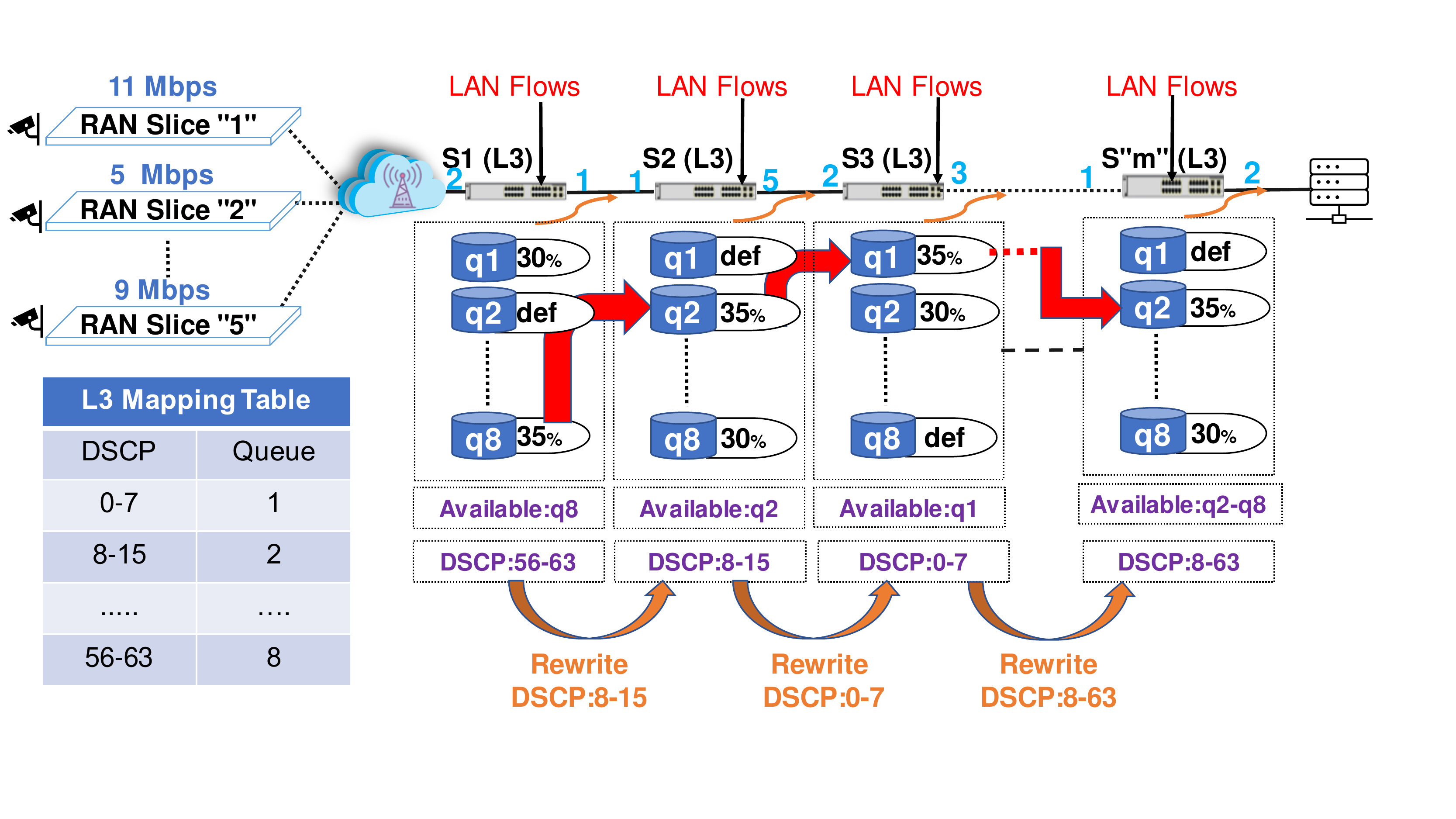}}
\caption{QoS mismatch in layer 3 switches}
\vspace{-0.2 in}
\label{fig:l3-mismatch}
\end{figure}


\subsubsection{QoS mismatch in layer three switches \label{sec:goal1}} 
Figure~\ref{fig:l3-mismatch} shows a part of the enterprise LAN with four-layer 3 (L3) switches. Five slices from the RAN trunk to switch S1 in the enterprise LAN. As shown in the figure, other non-RAN flows enter the L3 switches as well. The egress port of every switch has eight queues labelled {\it q1} through {\it q8}. Also, each L3 switch has a table that maps a range of DSCP identifiers to one of the eight queues. For example, packets with DSCP identifiers from 0-7 are sent to queue {\it q1}. The figure also shows how the 1Gbps egress bandwidth of a port is apportioned among the eight queues of the egress port. For example, queue  {\it q1} is apportioned  30\% of the egress port bandwidth. Similarly, queue {\it q2} is the {\it default}, which means that packets with no DSCP identifiers are put in this queue, and packets are sent out of the egress port on a best-effort basis.

We illustrate a QoS mismatch that can occur in a layer three switch.
As shown in the fig \ref{fig:l3-mismatch}, let us assume that the available queues (i.e., these queues can carry new packets without dropping any packets) are queue {\it q8} in switch S1 (DSCP range 56-63) that has an unused capacity of 35\%, queue {\it q2} in switch S2 (DSCP range 8-15), and queue {\it q1} in switch S3 (DSCP range 0-7). We assume that the rest of the queues in the figure are fully subscribed by other non-RAN flows (shown in the figure as LAN flows) using a DSCP marking of 39. When the RAN packets with a DSCP marking of 39 enter the LAN, they will be placed in queue {\it q5} in switch S1 (as per the mapping table shown in the figure for the switch S1), queue {\it q5} in switch S2, and queue {\it q5} in switch S3. These RAN flows must, without a doubt, compete with existing LAN data in queue {\it q5} in switches S1, S2, and S3. Such competition usually breaks the Slice QoS requirements for the RAN flows, and packets from the RAN may be dropped.

In contrast, if the RAN flow was assigned a DSCP marking in the range 56-63 for switch S1, a DSCP marking in the range 8-15 for switch S2, and a DSCP marking in the range 0-7 for switch S3, then the RAN flow would have no contention with the existing LAN flows that are using the DSCP marking of 39. A similar mismatch happened to RAN flows in the motivational experiment discussed in section \ref{sec:motivation}, where RAN flows had DSCP marking of 39. 


\begin{figure}[!t]
\vspace{-0.15 in}
\centerline{\includegraphics[width=1\linewidth]{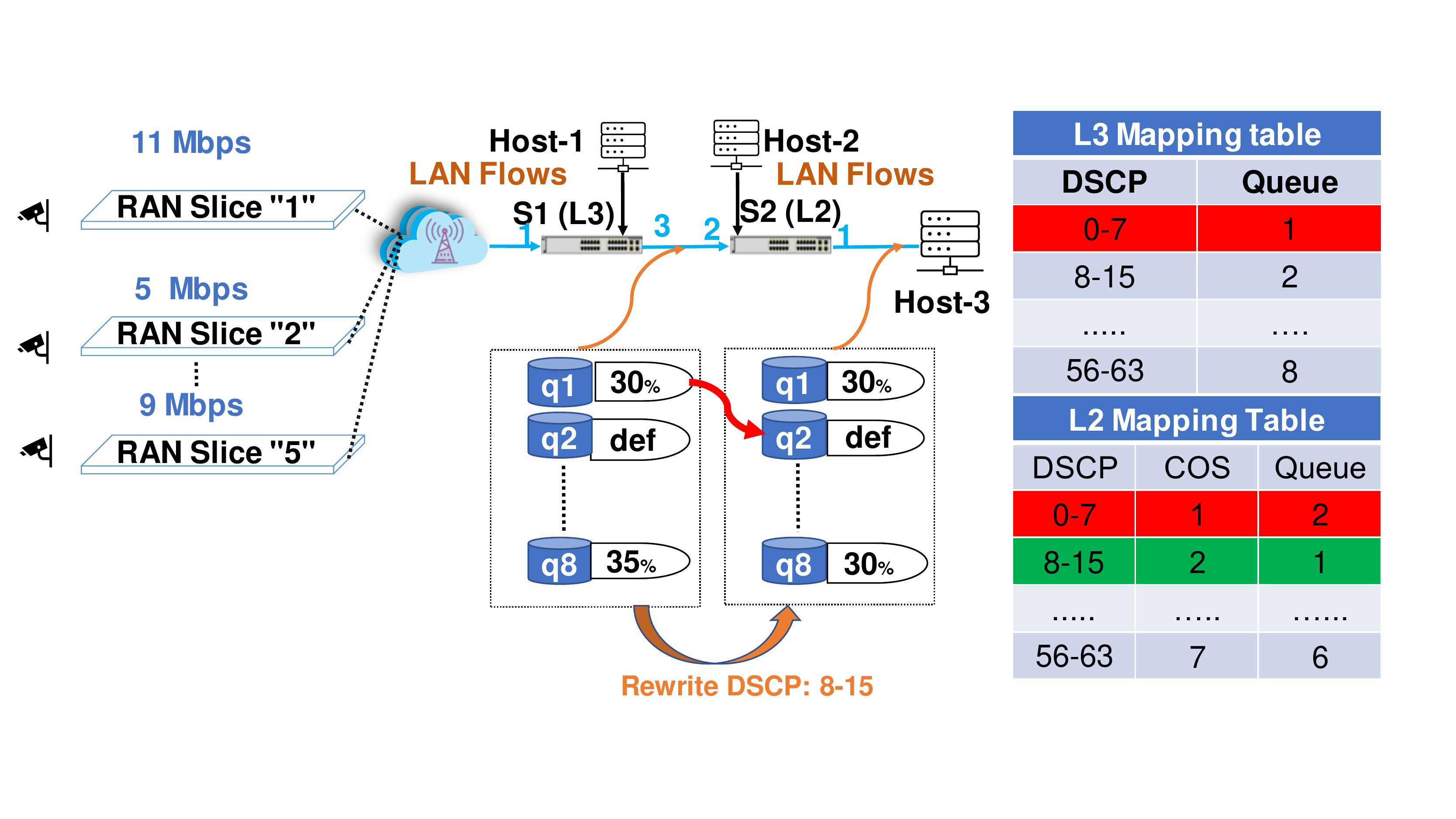}}
\caption{QoS mismatch between layer 3 and layer 2 switches}
\vspace{-0.2 in}
\label{fig:l2-l3-mismatch}
\end{figure}

\subsubsection{QoS mismatch due to layer 3 and layer 2 switches \label{sec:goal2}}
Enterprise LANs have both layer two (L2) and L3 switches. Unlike L3 switches, an L2 switch uses the 3-bit CoS (Class of Service) field in the ethernet header to assign frames to queues at an egress port. If the packets have a DSCP marking, then two translations occur: first, a DSCP to CoS translation is done, which is followed by a translation of CoS to queue number. This mapping table is shown as the L2 mapping table in Figure~\ref{fig:l2-l3-mismatch}. Here, packets with DSCP marking in the range 0-7 are mapped by an L3 switch to queue {\it q1} while the L2 switch maps the same DSCP range to a different queue. In the L2 switch, the DSCP range 0-7 translates to a CoS value of 1, which is mapped to queue {\it q2}. As shown in the figure, queue {\it q2} in an L2 switch is the default queue (best-effort traffic), and the Slice QoS can break when RAN flows traverse queue {\it q2} of the L2 switch. 

To preserve the slice QoS of the RAN flows, 5G data must use the DSCP range 0-7 to traverse the S1(L3) switch but change to the DSCP range 8-15 to traverse the S2(L2) switch so that the packets are assigned to queue {\it q1} in switch S2. Such an opportunistic change of DSCP identifiers can preserve slice QoS when 5G data traverses an L2 switch.


\subsubsection{QoS mismatch due to poor ingress-egress policies \label{sec:goal3}}
Packets received at ingress ports are put in an internal ring inside the switch.  When the total inbound bandwidth of all ports exceeds the bandwidth of the internal ring~\cite{internal-ring}, then packets are dropped before they are even put in ingress queues (ingress queues are located after the packet is classified, policed, and marked in the internal ring, and before packets are forwarded into the switch fabric).
A similar problem occurred for RAN flows in the motivation experiment in section \ref{sec:motivation}, where the LAN flows had too many packets, and some packets in the RAN flows were dropped at the internal ring. To mitigate this, policies had to be set at ingress port for RAN flows to reserve adequate bandwidth on the internal ring. This ensured that RAN flows got ingress priority over LAN flows.

At the egress port of an enterprise switch, packets are drained from the queues based on the queue's priority and the queue number. The higher the queue number, the higher the priority. There are two policies in LAN switches that control the queue draining: strict and round-robin. Strict policy continuously drains the packets from higher priority queues unless they are empty. Round-robin drains the packets in a round-robin fashion based on the maximum bandwidth allocated to the queues. A similar problem occurred for RAN flows in the motivation experiment in section \ref{sec:motivation}, and the egress queue draining policy had to be set to round-robin to preserve the QoS of the RAN flow.


\subsection{Challenges:}Design of a solution like 5GLoR poses significant challenges:
\begin{itemize}
    \item Topology Challenge: For 5GLoR to check the queue availability, mapping tables, and priority of ingress and egress ports of different switches, knowledge of the network path (switches and port details) from the IoT sensors to the LAN-resident application is essential. 
    \item Feasibility Challenge: To determine a possible path through the queues of different switches, knowledge of the status of the different queues (i.e., max capacity and current usage) is essential. 
    \item Adaptability Challenge: 5GLoR should be able to automatically adapt as new switches are added to the network path between the IoT sensors and the application.
    \item Heterogeneity Challenge: The QoS mechanisms and policies are the same on different switches from different vendors. However, the APIs used to invoke the mechanisms and policies are quite different across the switches. 5GLoR must work seamlessly with switches from vendors and SDN controllers that manage many L3 and L2 switches.  
\end{itemize}

\subsection{Key tasks and components in 5GLoR}\label{subsec:algorithm}

\begin{algorithm}[t]
\caption{Path discovery from sensor on RAN to application on LAN}
\label{algo:algo1new}
\begin{algorithmic}[1]
  \STATE \textbf{Input:}\\ 
  $IP_r$: IP address of sensor in RAN, $IP_l$: IP address of application in LAN,$R_{init}= $ router connecting RAN to LAN
  \STATE \textbf{Output:}\\
  \begin{math}
  P =\{(R_{init},p_{in},p_{out}),...,(R_{end},p_{in},p_{out})\}
  \end{math}
  \STATE $R_i$=$R_{init}$
  \WHILE{true}
  \IF{$IP_l$ is in same subnet as $IP_r$}\label{algo:arp}
    \STATE Get egress port $p_{out}$ on $R_i$ to reach $IP_l$;
    \STATE Append $\{R_i,,p_{out}\}$ to path $P$;
    \STATE {\bf break} \label{algo:endforwardpath}
   \ENDIF
   \IF{$IP_l$ is in same subnet as $R_i$}
    \STATE Get egress port $p_{out}$ on $R_i$ to reach $IP_l$;
    \STATE Append $\{R_i,,p_{out}\}$ to path $P$;
    \STATE {\bf break} \label{algo:endforwardpath}
    \ENDIF

     \STATE Determine next-hop from IP table of $R_i$;
     \STATE Get egress port $p_{out}$ on $R_i$ to reach next-hop;
    \STATE Append $\{R_i,,p_{out}\}$ to path $P$;
    \STATE Assign $R_i$ = next-hop ;
  \ENDWHILE
\end{algorithmic}
\end{algorithm}


\subsubsection{\textbf{Path Discovery}}
Path Manager finds the sequence of network switches (ingress and egress ports on each switch) from a sensor on RAN to the application on the VLAN. Utilities like traceroute \cite{b2} only focuses on routers on the path, but we also need details about the layer 2 switches along the path. Furthermore, utilities like traceroute do not provide (ingress and egress) port-level details. Therefore, we designed an algorithm that consults the MAC, ARP and IP tables on a layer 3 switch, and the MAC table in a layer 2 switch to determine the path with port-level details.  If all the switches are managed by an SDN controller, we can also get the path by querying the controller \cite{openflow}. In this paper, we assume that all switches are IP-accessible either directly or through an SDN controller, and the IP address of the first network element (layer 3 or layer 2 switch) that connects the RAN to the LAN is known. 

Path Manager implements the algorithm shown in Figure~\ref{fig:sys} to determine the layer 3 switches (and ingress or egress ports) on the path. Since a layer 3 switch is also a layer 2 switch, we can consult the MAC, ARP and IP tables to determine the ingress and egress ports on each switch. We begin with the network element that connects the RAN to the LAN ($R_{init}$). If the destination (i.e. the application on the VLAN) is on the same subnet as the source (i.e. the sensor on the RAN), then ARP (or ARP cache) provides the MAC address of the destination, and the MAC table has the egress port, and the forward pass of the path discovery process is terminated. Otherwise, we check if $R_{init}$ is on the same subnet as the destination. If yes, then we use ARP and the MAC table to determine the egress port on $R_{init}$ to reach the destination, and the forward pass terminates. Otherwise, we lookup the IP table for the next hop, determine the MAC address of the next hop, and find the egress port on $R_{init}$ to reach the next hop. This information is added to the path, and we repeat the above process for the next hop $R_i$. 

The forward pass shown in Figure~\ref{fig:sys} determines the layer 3 switches on the path, as well as the egress ports on all these switches. To determine the ingress ports on these switches, we use a similar algorithm as in Figure~\ref{fig:sys}, but this time we process the switches (in the path) in a reverse order (i.e. from the application on LAN to the sensor on the RAN).

The algorithm shown in Figure~\ref{fig:sys} does not identify the layer 2 switches on the path. Therefore, our implementation uses a variant of traceroute~\cite{layer2-traceroute} that can be used to find all the layer 2 switches between any two routers on the path. Since we assume that all switches are accessible using their IP address, we can consult their MAC tables to determine the ingress and egress ports. 

Alternatively, the network path can also be specified manually and 5GLoR can determine the ingress and egress ports automatically along the given path.


\subsubsection{\textbf{Feasibility Check}}:
We carefully examine each switch on the path to check if it can accomodate a given RAN flow. Our objective is to determine the egress queue that can support the flow, and the DSCP identifier that should be used to traverse this switch.

To check feasibility at a particular switch, we have to estimate the LAN flows in each of the egress queues. By querying the switch, we can determine the configured, maximum capacity of each egress queue. If we also know the LAN flow into the egress queue, then it is easy to check if the given RAN flow can be supported (without dropping any packets) by this egress queue.

We periodically estimate the LAN flow into each egress queue by using sFlow~\cite{b5}. 
It is a multi-vendor, packet sampling technology that 
samples 1 in N packets passing through the interface, irrespective of the flow. 5GLoR uses the ToS (type of service) field from the sFlow summary to estimate the number of ToS-based flows. ToS has the DSCP marking that allows us to infer the queue for the packet. 


We checks the status of every switch in the path. At each switch, we examine all the queues for the egress port. If there is a queue with adequate capacity to carry the RAN flow, then the procedure chooses a DSCP identifier that will map the RAN flow into the queue. Information about the LAN flows in this queue are pre-computed by using sFlow. By querying the switch, we also know the pre-set capacity limit for the queue. Since we know the desired RAN flow, the difference between pre-set capacity and the existing LAN flow is sufficient to determine whether the queue can support the new RAN flow. It is possible that we may not find a suitable DSCP identifier for a switch in the path. Then, this switch may break the QoS of the RAN flow. The feasibility check procedure is implemented in the feasibility manager shown in Figure~\ref{fig:sys}.

\begin{figure}[t]
    \centering
    \begin{subfigure}{0.45\columnwidth}
        \centering
        \includegraphics[width=\columnwidth,height=0.9 in]{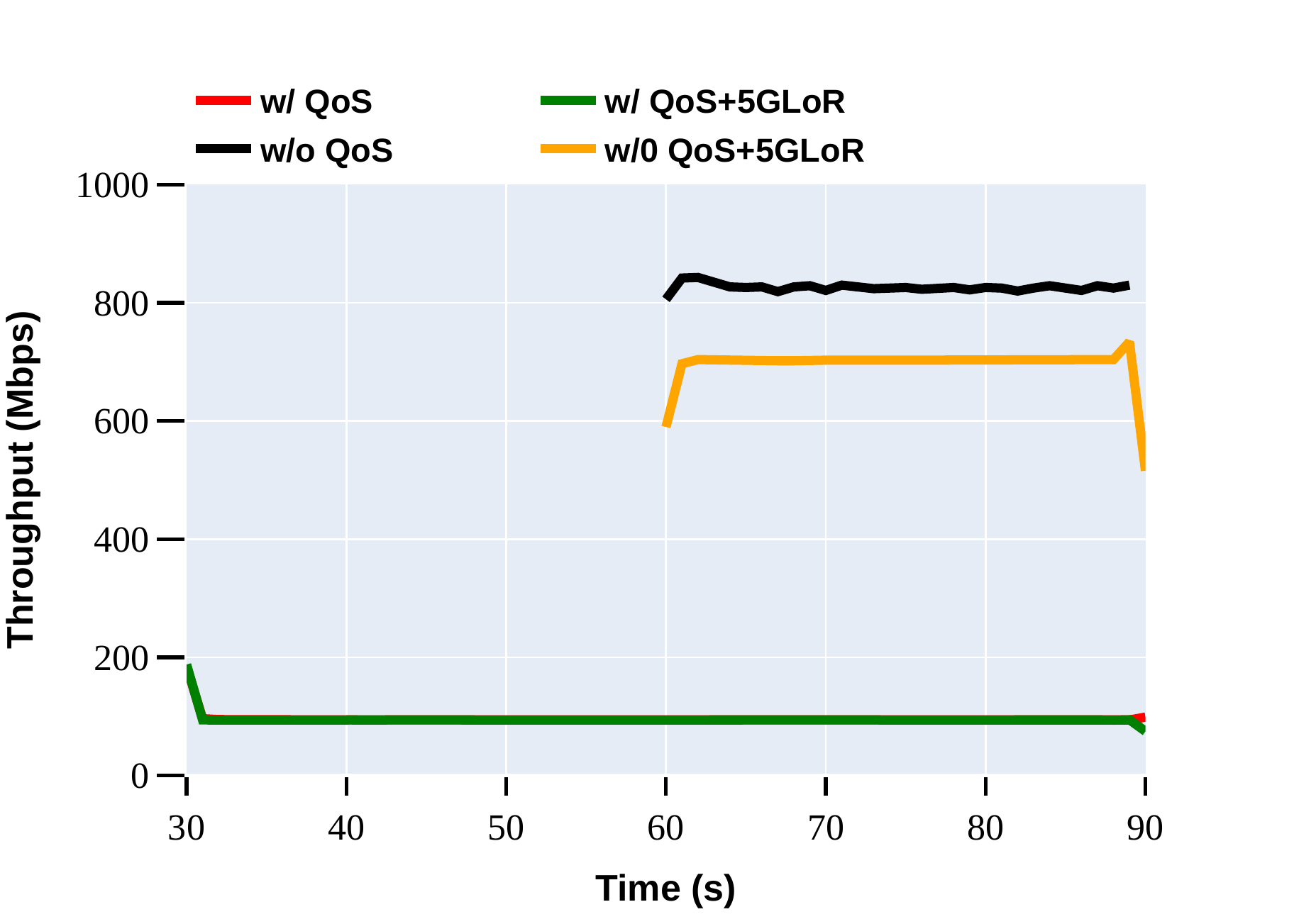}
        \vspace{-0.2 in}
        \caption{Throughput}\label{fig:exp-tput-lanqos}
        
    \end{subfigure}
    \begin{subfigure}{0.45\columnwidth}
        \centering
        \includegraphics[width=\columnwidth,height=0.9 in]{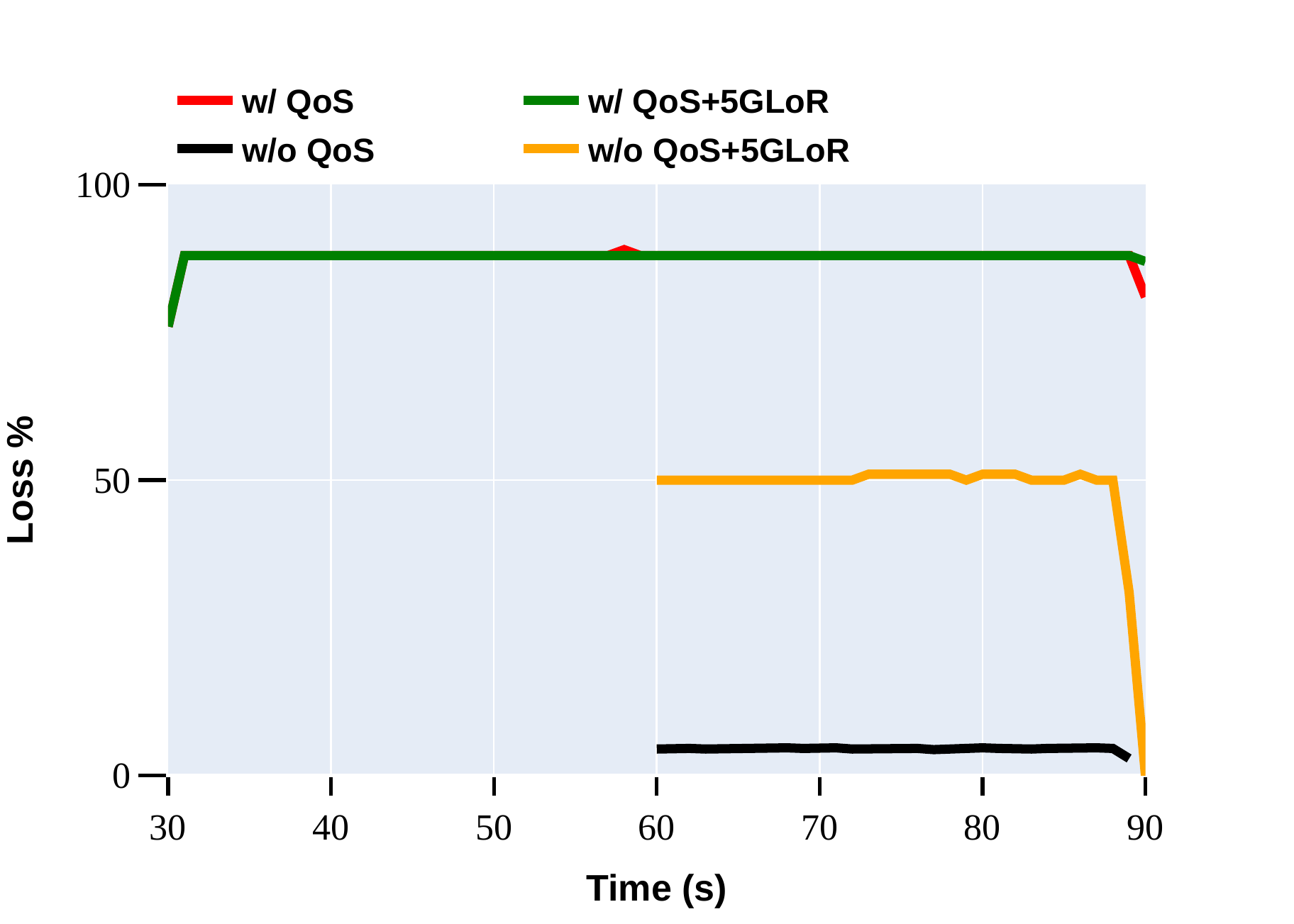}
        \vspace{-0.2 in}
        \caption{Loss}\label{fig:exp-loss-lanqos}
    \end{subfigure}
   \caption{LAN Flows with 5GLoR with right DSCP}
    \label{fig:exp-lanflows}
    \vspace{-0.1 in}
\end{figure}

\begin{figure}[t]
    \centering
   \begin{subfigure}{0.45\columnwidth}
       \centering
       \includegraphics[width=\columnwidth,height=0.9 in]{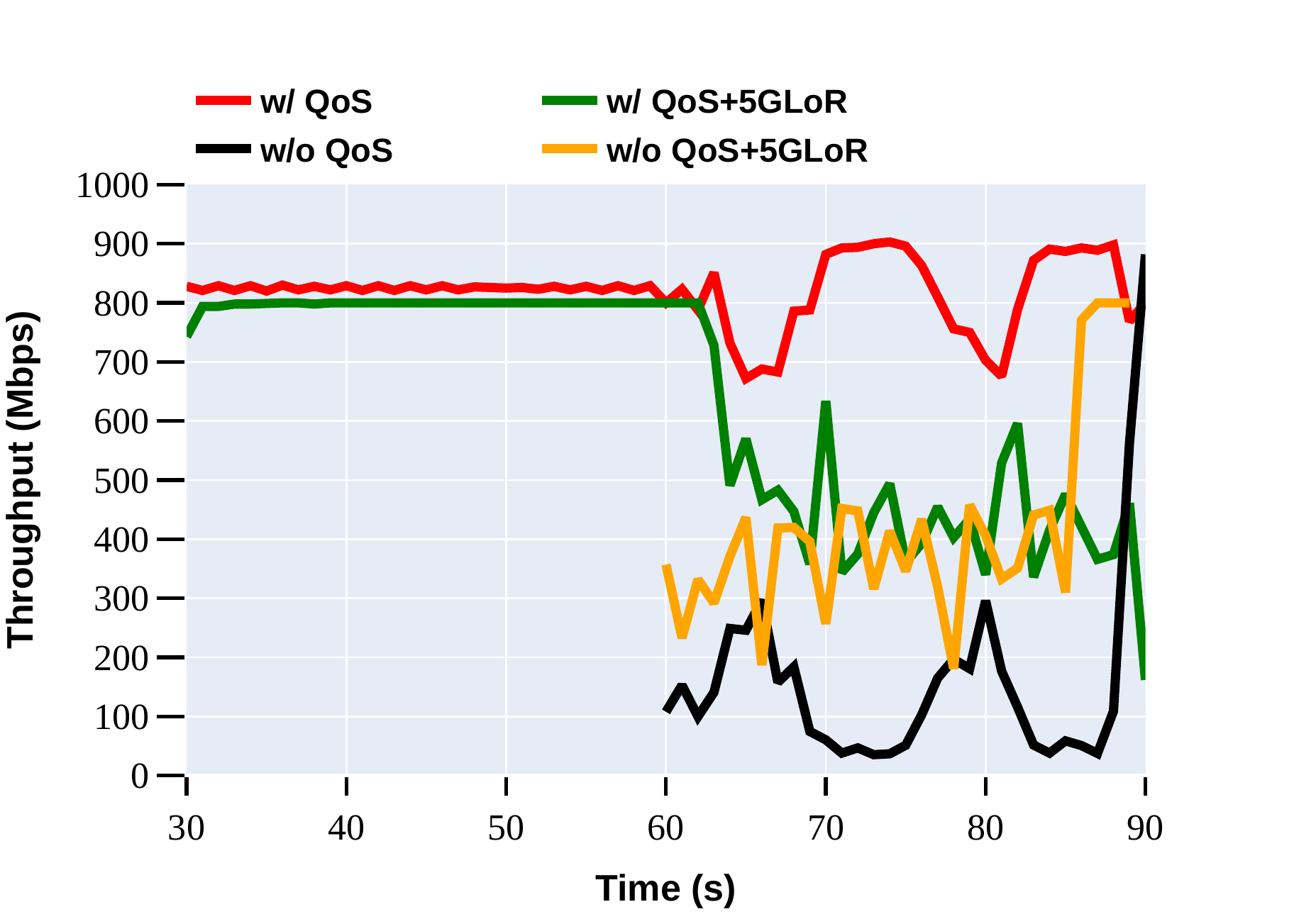}
       \vspace{-0.2 in}
       \caption{Throughput}\label{fig:exp-tput-lan}
   \end{subfigure}
   \begin{subfigure}{0.45\columnwidth}
       \centering
       \includegraphics[width=\columnwidth,height=0.9 in]{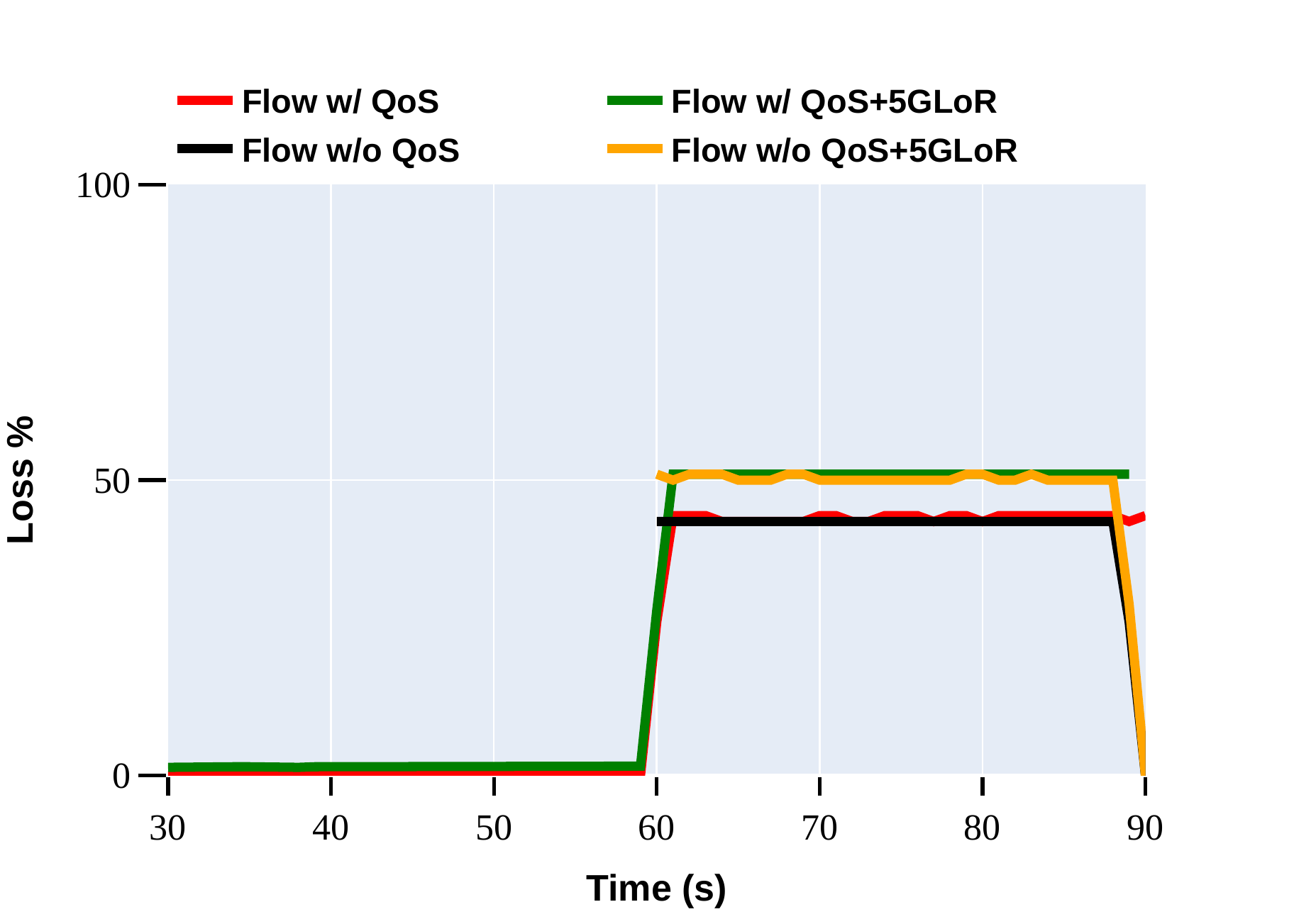}
       \vspace{-0.2 in}
       \caption{Loss}\label{fig:exp-loss-lan}
   \end{subfigure}
        
   \caption{LAN Flows with 5GLoR with default Queue DSCP}
    \label{fig:exp-lanflows-defaultqos}
    \vspace{-0.1 in}
\end{figure}

\begin{figure*}[t]
    \begin{subfigure}{0.4\columnwidth}
        \centering
        \vspace{-0.08 in}
        \includegraphics[width=1\columnwidth,height=0.8 in]{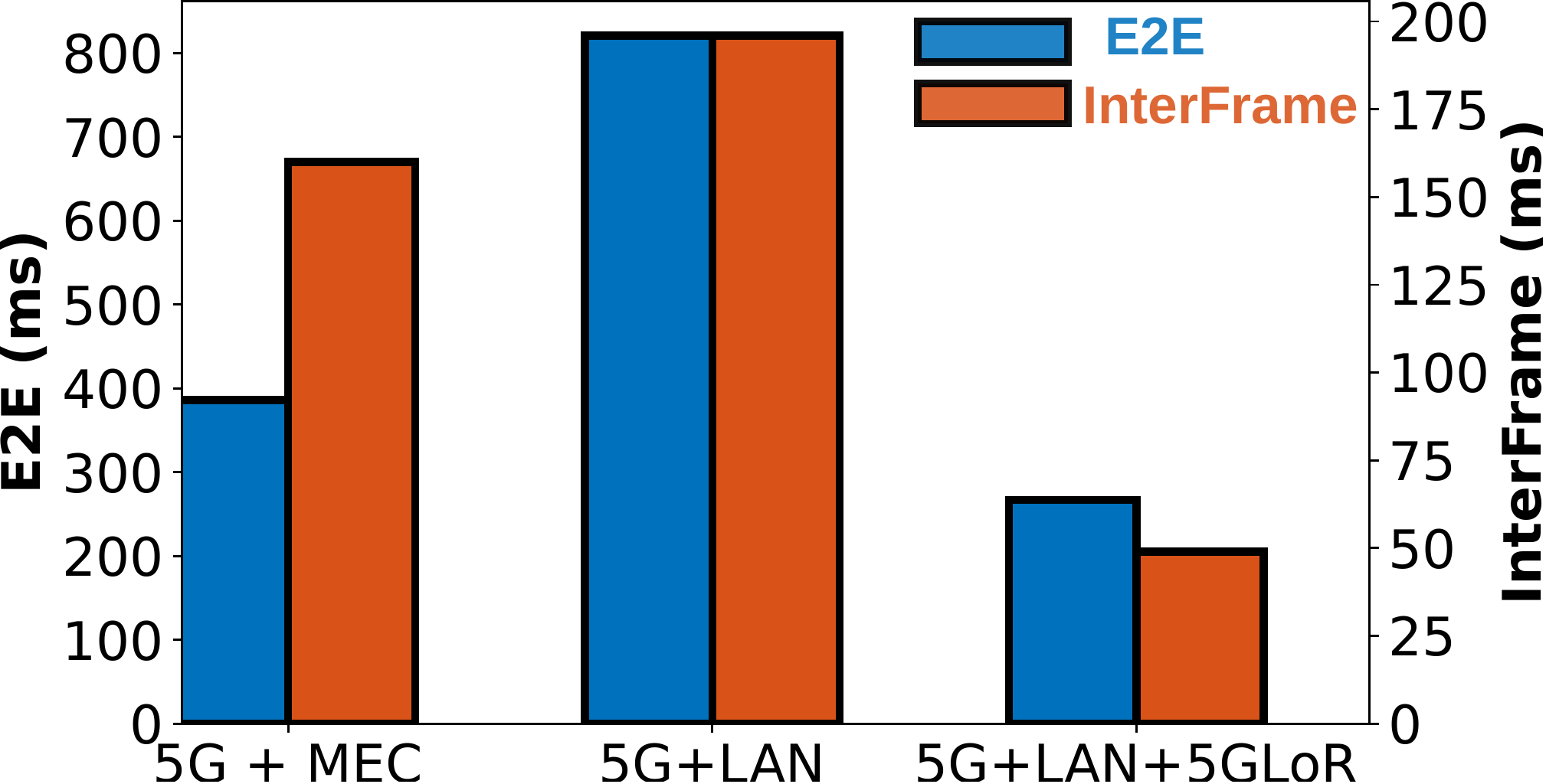}
        \vspace{-0.2 in}
        \caption{ E2E\& InterFrame }\label{fig:exp-e2eif}
    \end{subfigure}
    \begin{subfigure}{0.4\columnwidth}
        \centering
        \vspace{-0.08 in}
        \includegraphics[width=1\columnwidth,height=0.8 in]{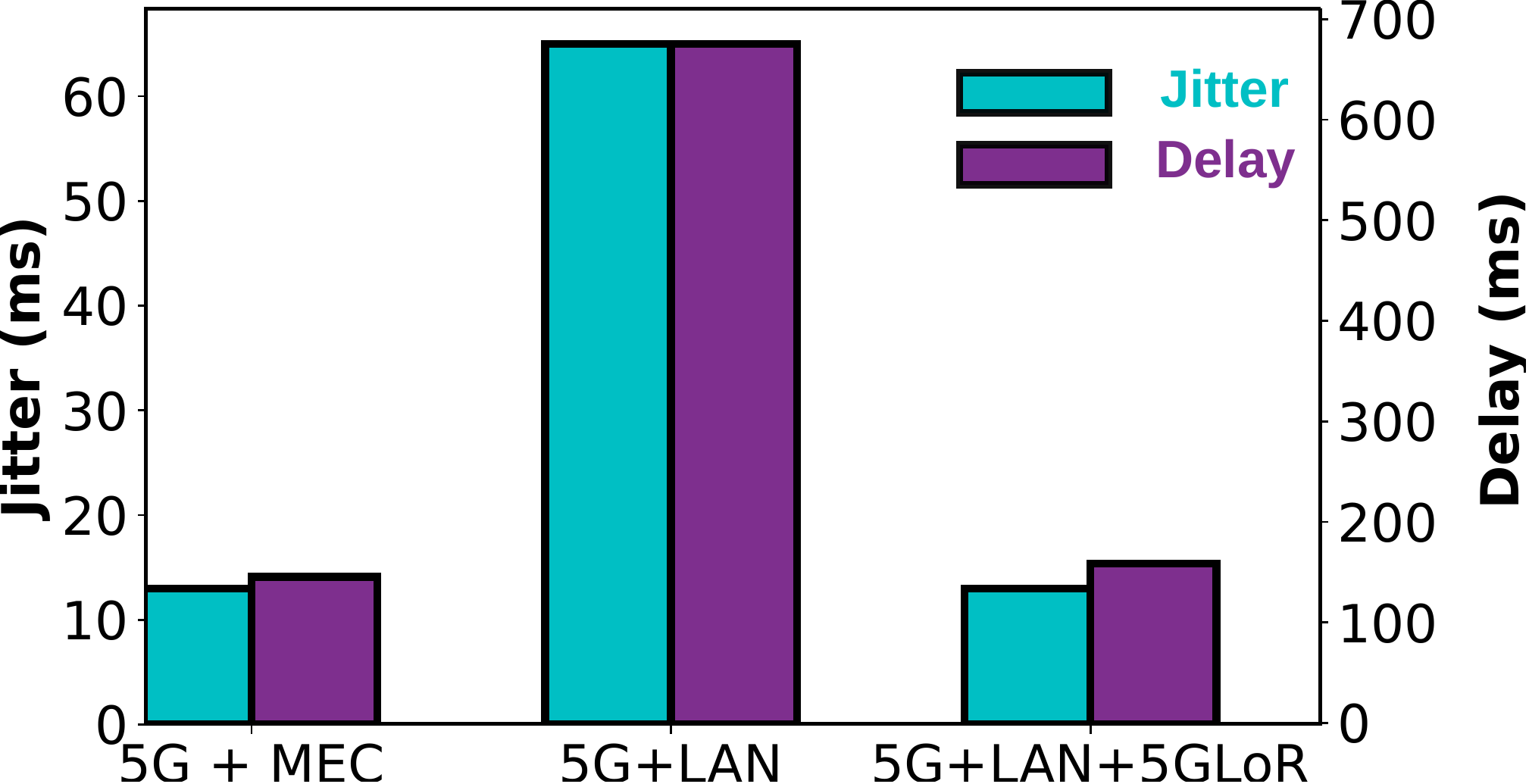}
        \vspace{-0.2 in}
        \caption{Jitter \& Delay}\label{fig:exp-jitter}
    \end{subfigure}
    \begin{subfigure}{0.4\columnwidth}
        \centering
         \vspace{-0.08 in}
        \includegraphics[width=1\columnwidth,height=0.8 in]{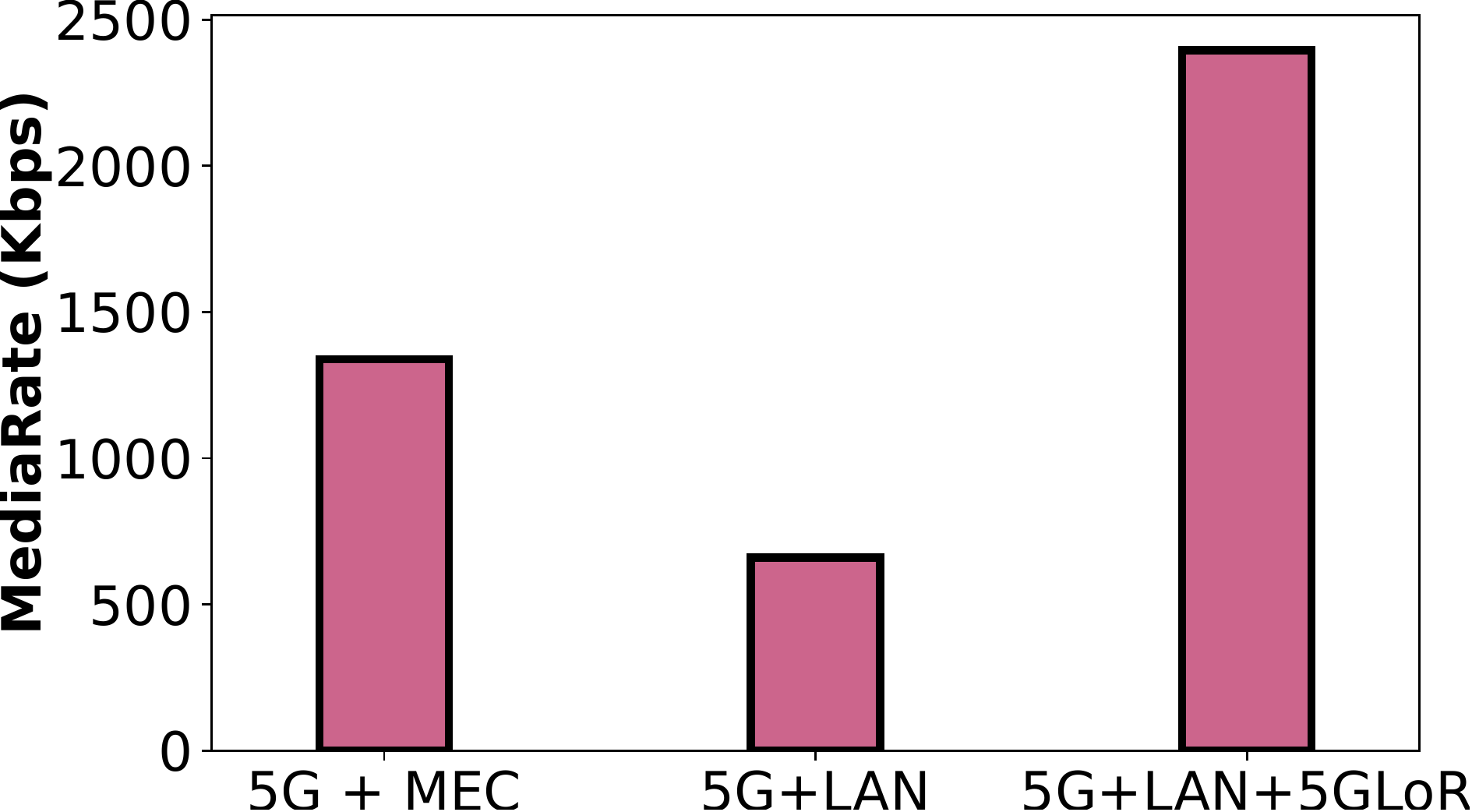}
        \vspace{-0.2 in}
        \caption{Media Rate}\label{fig:exp-mediarate}
    \end{subfigure}
    \begin{subfigure}{0.4\columnwidth}
        \centering
        \vspace{-0.08 in}
        \includegraphics[width=1\columnwidth,height=0.8 in]{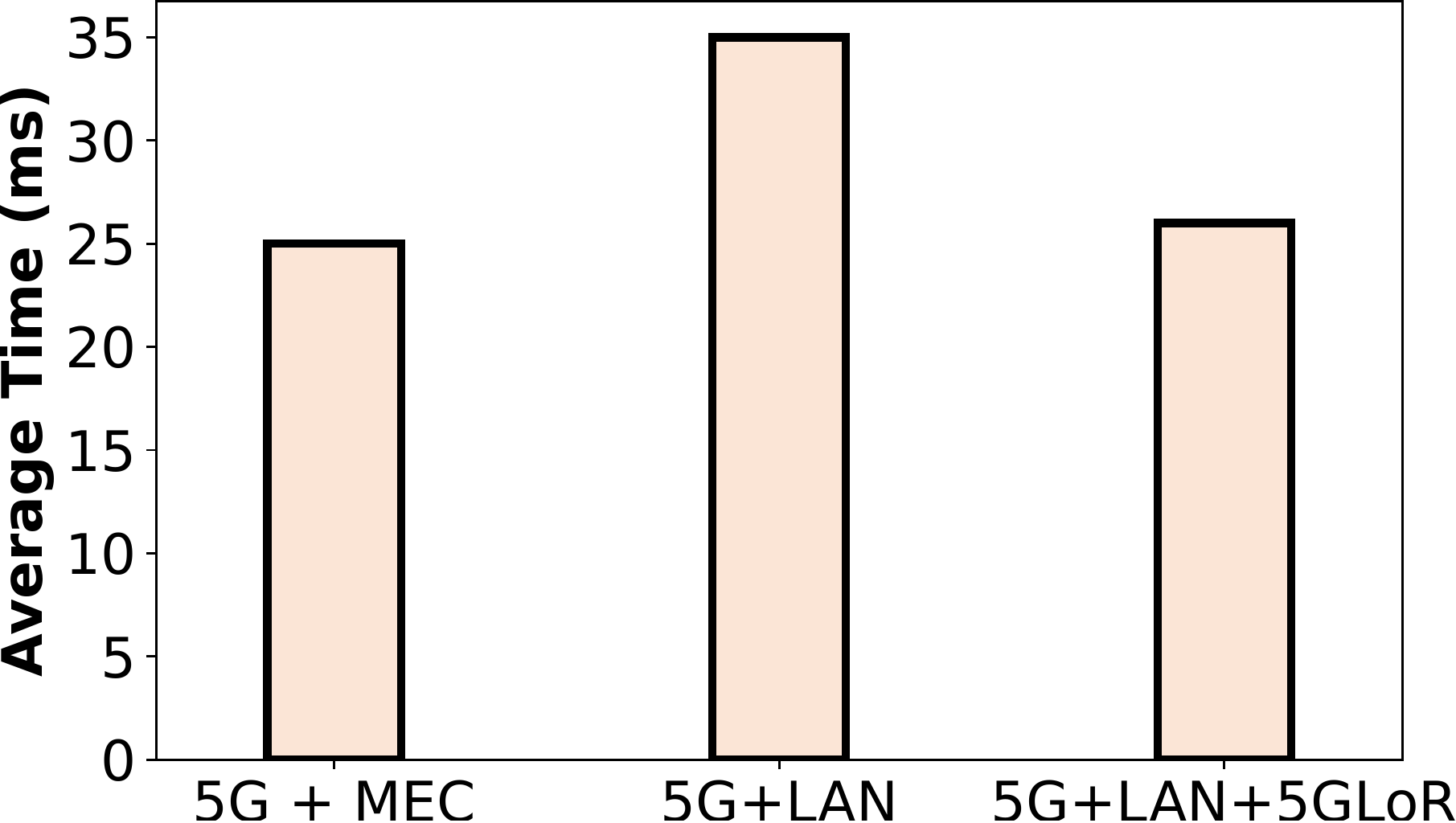}
        \vspace{-0.2 in}
        \caption{Detection Time }\label{fig:exp-fdtime}
    \end{subfigure}
    \begin{subfigure}{0.4\columnwidth}
        \centering
        \vspace{-0.08in}
        \includegraphics[width=1\columnwidth,height=0.8 in]{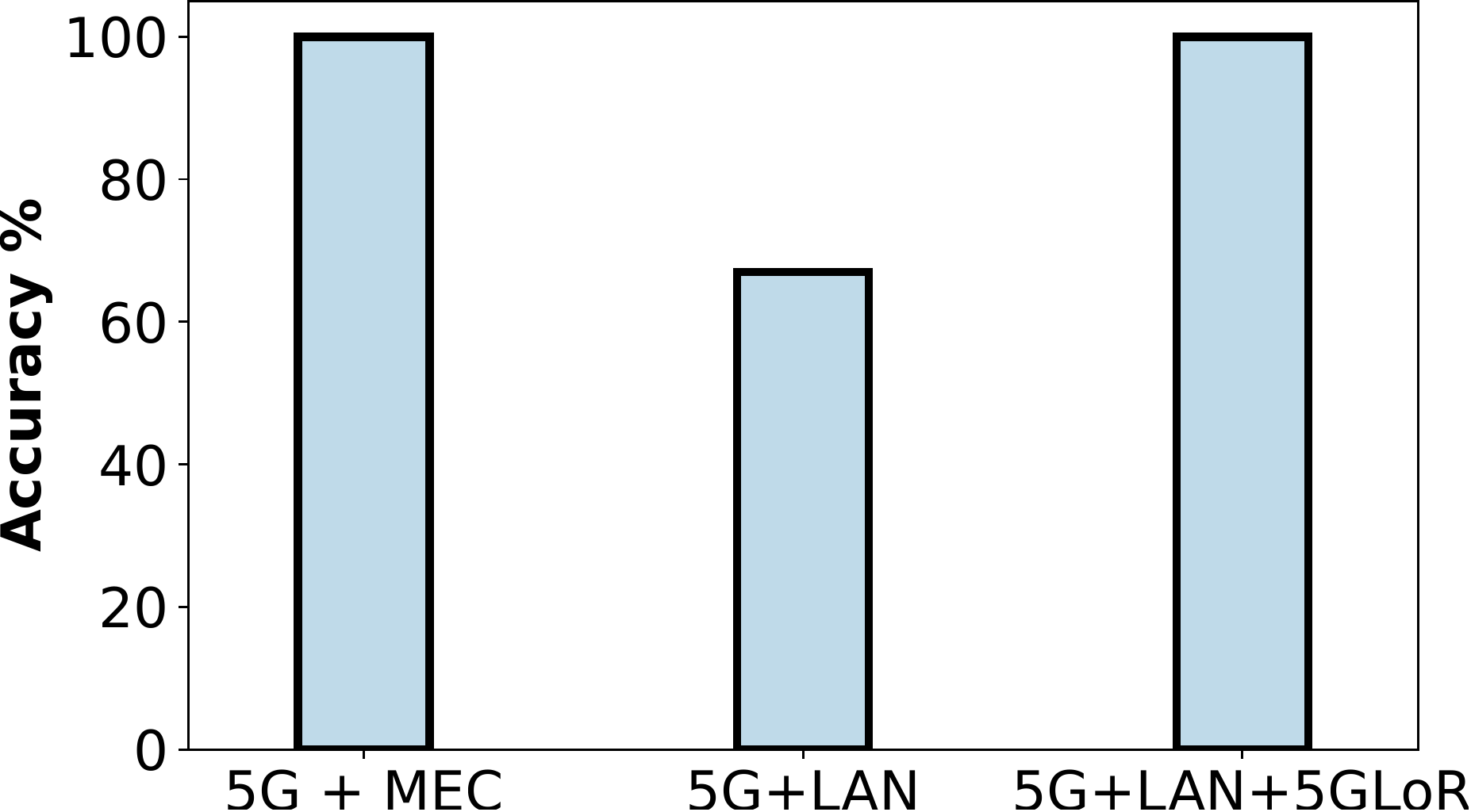}
        \vspace{-0.2 in}
        \caption{Detection Accuracy }\label{fig:exp-fdaccuracy}
    \end{subfigure}
    \caption{Comparison of 5G LAN, 5GLoR and 5G+MEC}
    \vspace{-0.2 in}
    \label{fig:exp-comp}
\end{figure*}

\subsubsection{\textbf{DSCP re-write rules}}
As explained in section \ref{sec:goal1}, it is possible that the Feasibility Check procedure may recommend a different queue number across different switches on the path. Therefore, the DSCP marking necessary to safely traverse a switch can be different across switches. To address this DSCP identifier mismatch across switches,  5GLoR installs rewrites rules \cite{b3} on appropriate switches so that the RAN flow is directed to the appropriate egress queue in each switch of the path. The DSCP identifier to be used for the first network element in the path is communicated to the application, and all packets from the application have this DSCP marking.
The Queue manager in Figure~\ref{fig:sys} installs the re-write rules on switches.

\subsubsection{\textbf{Ingress and egress policies}}
As explained in section \ref{sec:goal1}, inbound flows on different ingress ports can interfere.
To avoid this problem, we use the ingress reservation for ingress ports of all the switches in the path: we use the two-rate three-color technique. Two rate defines the min and maximum rate, and three colors, red, green, and yellow, decides the packet drop \cite{b4}. We define the $min_{rate}=min(\{gbr1,gbr2,...,gbr_f\})$ and $max_{rate}=max(\{gbr1,gbr2,...,gbr_f\})$. 
We also ensure that the queue draining policy is Round-Robin at egress ports of all the switches in the path.

\begin{figure}[!t]
\centerline{\includegraphics[width=1\linewidth]{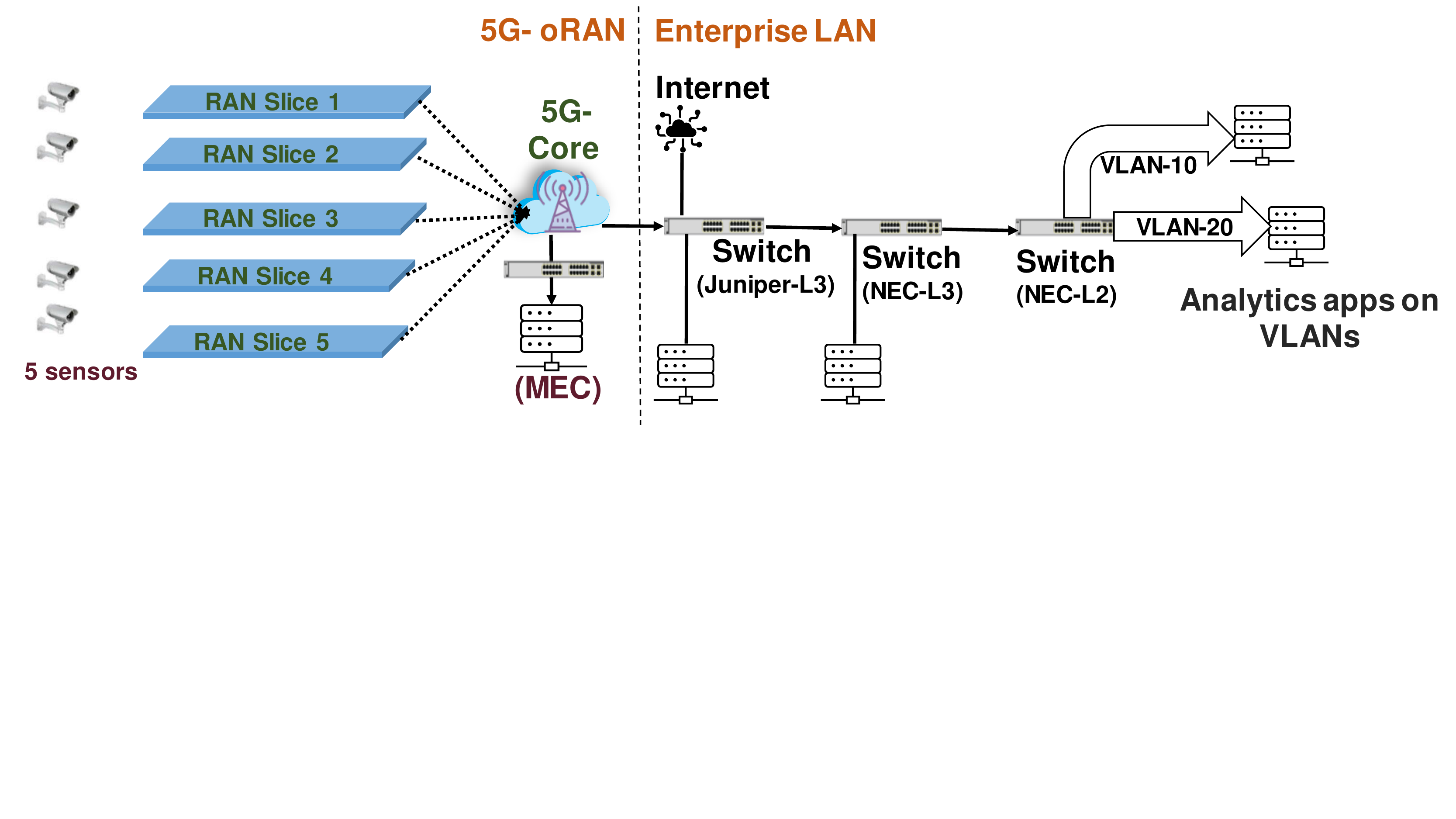}}
\vspace{-0.9 in}
\caption{Test-Bed: Block Diagram}
\label{fig:testbed-block}
\vspace{-0.2 in}
\end{figure}

\section{Evaluation}\label{sec:eval}
In this section, first, we explain the end-to-end test bed and then show the detailed evaluation of the 5GLoR.

\subsection{Our 5G LAN}
Key components of our 5G LAN are shown in Figure~\ref{fig:testbed-block}.
The 5G cellular network components (RAN and core) are from a commercial vendor \cite{celona}. Our UEs (user equipment) are from also from another commercial vendor \cite{multitech}. We have five five laptops (with Ubuntu 20 OS), and each laptop uses a distinct UE to stream stored videos over the 5G cellular network.  We can also connect an IP camera to each laptop to live stream video over the 5G network.  For our experiments, we use a public benchmark -- video of the publicly available airport security recording~\cite{pond5}. 

As shown in Figure~\ref{fig:testbed-block}, we also have one layer 3 ex2300 Juniper switch, two NEC switches (a layer 3 switch qx5800, and a layer 2 switch qx5200), and an SDN controller.  All the switches are accessed and controlled using APIs from Mist UI for Juniper and NOE UI (SDN controller) for NEC. Our MEC layer of compute is a laptop that is connected to the 5G core via a layer 2 switch. For our 5G LAN, the Juniper switch is the internet gateway. We also introduce additional LAN flows using servers that are directly connected to the Juniper and NEC switches. We also have two virtual LANs (VLANs) that host our applications. 

\subsection{Experimental Results}

\subsubsection{Impact of 5G LoR on LAN flows}\label{sec:lanflows}


In the motivation section (\ref{sec:motivation}), we showed that 
5GLoR ensured that the QoS of the RAN flows in both the second window (30-60 sec) and the third window (60-90 secs) are preserved, in spite of the additional DSCP marked and unmarked LAN flows.  

We now analyze the impact of 5GLoR and the RAN flows on the LAN flows.

The DSCP marked LAN flows map to egress queue 6 in switch S1. 
This queue has a pre-set capacity of 100 Mbps. However, the total incoming DSCP marked LAN flows is 1125 Mbps (75 x 15 Mbps). Therefore, we observe only 100 Mbps of output at the egress port. This is shown in Figure~\ref{fig:exp-tput-lanqos} as the green waveform. The DSCP marked LAN flows also experience high packet loss (green waveform in Figure~\ref{fig:exp-loss-lanqos}). Since the RAN flows are using egress queue 8 (with a pre-set queue capacity of 350 Mbps, we observe that 5GLoR has no impact on the DSCP marked LAN flows.
%

However, 5GLoR has a significant impact on the best-effort flows. Without 5GLoR, the 75 marked flows (with aggregate flow of 1500 Mbps) are put in the best-effort egress queue. Since most of the higher numbered queues (except queue 6 which has 100 Mbps of the marked LAN flows) are empty, the unmarked LAN flows egress the switch at almost 800 Mbps (red waveform in Figure~\ref{fig:exp-tput-lan}; egress port has a bandwidth of 1000 Mbps)! The packet loss for the unmarked LAN flows is small (red waveform in Figure~\ref{fig:exp-loss-lan}). However, when 5GLoR introduces a total of 40 Mbps RAN traffic  into queue 8 (five RAN flows:  6+6+6+11+11 Mbps), the throughput of the 75 marked LAN flows (best-effort flows) drops because of the higher-priority RAN flows. Consequently, the loss for the marked LAN flows is higher, as shown by the green waveform in Figure~\ref{fig:exp-loss-lan}.

\subsubsection{Impact of 5GLoR on enterprise IoT applications}
We consider three applications (face detection, face recognition, and WebRTC) to analyze the impact of 5GLoR. For each application, we create five RAN flows (flow 1 thru flow 5), and  encapsulate each flow in a separate network slice. We consider three scenarios: a 5G LAN, the 5G LAN augmented with the proposed technique 5GLoR, and 5G+MEC. For 5G LAN experiments, all our applications are hosted on the enterprise VLANs. For the 5G+MEC experiments, all our five applications are hosted on the MEC (i.e. computing resources at the edge of the 5G network; therefore, no traffic from the RAN enters the enterprise LAN). Figure~\ref{fig:exp-comp} shows results for all three scenarios.

{\bf Frame and packet delays:}
Figure~\ref{fig:exp-e2eif} shows the impact of 5GLoR on WebRTC~\cite{WebRTC} flows. WebRTC is a web real-time communication framework for browsers, and many popular applications like Facebook messenger, WhatsApp, and Snapchat use this communication framework. The average end to end (E2E) delay for a frame, as well as the average inter-frame delay for the WebRTC flows are decreased by 212\% and 122\%, respectively, by using 5GLoR. 

Compared to the 5G+MEC case, note that the average frame delay and inter-frame delay are reduced by 30\% and 70\% respectively (see Figure~\ref {fig:exp-e2eif}).  The reason for better performance compared to MEC is the usage of DSCP in WebRTC\cite{WebRTC}. We also analyzed the per-packet jitter and delay for the WebRTC streams.With 5GLoR, per-packet jitters and delays are reduced by 80\% and 67 \% (see Figure~\ref{fig:exp-jitter}) when compared to 5G LAN without 5GLoR. Furthermore, these delays are comparable to the delays observed for the 5G+MEC scenario. As shown in Figure~\ref{fig:exp-mediarate}, the media rate with 5GLoR is 2.5 Mbps, which is 3.6 times better than the 5G LAN scenario. The mediarate for the 5G+MEC scenario was 1.5Mpbs, which is much lower than the mediarate for the 5GLoR case.
 

{\bf Accuracy:}
Figures~\ref{fig:exp-fdtime} and~\ref{fig:exp-fdaccuracy} show the impact of 5GLoR on a face detection application, which detects human faces in a video frame by using machine learning techniques. As shown in Figure~\ref{fig:exp-fdtime}, the average time to detect a face improves by 25\% when 5GLoR is used, and this time is comparable to the average detection time for the 5G+MEC scenario. 

We also measured the accuracy of face detection for the three scenarios. As shown in Figure~\ref{fig:exp-fdaccuracy}, we observe much higher accuracy (33\% higher) when 5GLoR is used in the 5G LAN. Also, accuracy of face detection for the 5G LAN with 5GLoR scenario is comparable to the 5G+MEC scenario.

\section{Conclusion}
Although 5G technology has the unique ability to distinguish among customers in the same traffic class (by using nework slicing), and offer differentiated services to different customers, this ability is not supported by most enterprise LANs. As a consequence, applications resident on the enterprise LAN are unable to take advantage of the unique abilities of 5G. To address this gap, we proposed 5GLoR that helps enterprise IoT applications achieve predictable performance.
Our extensive experiments demonstrated that for a variety of figures of merit like throughput, delay, jitter, and accuracy, a 5G LAN with the proposed 5GLoR technique is superior to a 5G LAN without our technique. Furthermore, a 5G LAN with the proposed 5GLoR technique is very competitive to the alternative of hosting applications in 5G+MEC. This is significant because 5G LAN offers an order of magnitude more computing, networking, and storage resources to the applications than the resource-constrained MEC, and mature enterprise technologies can be used to deploy, manage and update enterprise IoT applications.


\end{document}